\documentclass[a4paper,fleqn,usenatbib]{mnras}

\usepackage[T1]{fontenc}
\usepackage{ae,aecompl}

\usepackage{graphicx}	
\usepackage{amsmath}	
\usepackage{amssymb}	

\title[Oscillating Universe]{Black Holes and Neutron Stars in an Oscillating Universe}

\author[N.N. Gorkavyi$^{1}$ and S. A. Tyul'bashev$^{2}$]{
N.N. Gorkavyi$^{1}$\thanks{E-mail: nick.gorkavyi@gmail.com}
and S. A. Tyul'bashev$^{2}$\thanks{E-mail: serg@prao.ru}
\\
$^{1}$Science Systems and Applications, Lanham, 20706 USA
\\
$^{2}$Lebedev Physical Institute, Pushchino Radio Astronomy Observatory,
Russian Academy of Sciences, Pushchino, 142290 Russia
}

\date{Accepted March 3, 2021. Received May 6, 2020}

\pubyear{2021}

\begin{document}
\label{firstpage}
\pagerange{\pageref{firstpage}--\pageref{lastpage}}
\maketitle

\begin{abstract}
In recent years, hypotheses about a cyclical Universe have been again
actively considered. In these cosmological theories, the Universe,
instead of a ``one-time'' infinite expansion, periodically shrinks
to a certain volume, and then again experiences the Big Bang. One of the problems
for the cyclic Universe will be its compatibility with a vast population
of indestructible black holes that accumulate from cycle to cycle.
The article considers a simple iterative model of the evolution of black holes
in a cyclic Universe, independent of specific cosmological theories.
The model has two free parameters that determine the iterative decrease
in the number of black holes and the increase in their individual mass.
It is shown that this model, with wide variations in the parameters, explains
the observed number of supermassive black holes at the centers of galaxies,
as well as the relationships between different classes of black holes.
The mechanism of accumulation of relict black holes during repeated pulsations
of the Universe may be responsible for the black hole population detected
by LIGO observations and probably responsible for the dark matter phenomenon.
The number of black holes of intermediate masses corresponds to the number
of globular clusters  and dwarf satellite galaxies. These results argue for models
of the oscillating Universe, and at the same time impose
substantial requirements on them.
Models of a pulsating Universe should be characterized by a high level
of relict gravitational radiation generated at the time of maximum compression
of the Universe and mass mergers of black holes, as well as solve
the problem of the existence of the largest black hole that is formed
during this merger.
It has been hypothesized that some neutron stars
can survive from past cycles of the Universe and contribute to dark matter.
These relict neutron stars will have a set of features by which they can be
distinguished from neutron stars born in the current cycle of the birth
of the Universe. The observational signs of relict neutron stars and
the possibility of their search in different wavelength ranges are discussed.
\end{abstract}




\section{Introduction}

Black holes are one of the main components of the Universe. It is well known
that Stellar Black Holes (SBH) with a mass less than $100M_{\odot}$ are formed
in the process of stellar evolution. The number of SBHs that appeared
in the Universe after the Big Bang is estimated to be about $0.1\%$
of the total number of stars $10^{23}$, which is $10^{20}$ SBHs \citep{Cherepashchuk2014}.

In his seminal book, the Nobel laureate \citet{Peebles1993} begins
the section on the oscillating Universe with the statement that before
the theory of inflation, the most popular cosmological model was the Lemaitre model
of the Phoenix Universe, which disintegrates after the collapse
of the previous phase of the periodic Universe. Note that for the first time
the question of the cyclic Universe within the framework of Einstein's theory
was considered by A. A. Friedman in his famous work of 1922. The period from
1922 to the 1980s can be called the time of the dominance of the classical model
of the oscillating Universe. Despite the great fascination of recent decades
with the quantum theory of a one-time universe, oscillating cosmological models
are now again attracting a lot of attention
\citep[see reviews of][]{Brandenberger2017, Novello2008}.  Note the periodic model of another
Nobel laureate \citet{Penrose2011}, as well as the theory of \citet{Steinhardt2002}.
But models of the oscillating Universe rarely take into account the existence
of a large population of black holes. But in every cycle of the Universe,
there is a constant influx of black holes due to supernova explosions.
These black holes are indestructible and must accumulate from cycle to cycle,
as well as entropy (supermassive black holes, in fact, are the main carriers
of the entropy of the Universe). Therefore, any cyclic model of the Universe
must be accompanied by a model of the evolution of black holes.
This article is devoted to a simple model of the evolution of black holes
and neutron stars in the classical model of the oscillating Universe.
Note that the study of the population of black holes has become
a particularly relevant topic in recent years.

In 2015--2019, the LIGO gravitational wave observatory detected a large population
of black holes 8--80$M_{\odot}$, with a typical mass of $30M_{\odot}$ \citep{Abbot2016}.
It was immediately suggested that these numerous SBHs could explain
the dark matter phenomenon \citep{Bird2016, Kashlinsky2016}. The articles
\citet{Clesse2017, Clesse2018, Garsia2018} summarize the arguments in favor of such a hypothesis
and prove that LIGO, due to observational selection, records only the heaviest
black holes of stellar masses, and in fact the maximum number of black holes
should fall on masses of several solar masses. \citet{Carr2018} also believe
that dark matter can consist entirely of black holes. The hypothesis
that explains the phenomenon of dark matter using SBHs is attractive,
but it assumes a large number of such black holes---$10^{23}$, which is
three orders of magnitude more than the theories of stellar evolution suggest.
According to other authors, black holes can only make up a fraction
of the total mass of dark matter, but even if they are only about $ 10\%$,
this number of SBHs is two orders of magnitude higher than the theoretical estimate.

At the center of each galaxy are supermassive black holes (SMBHs) with masses of
\mbox{$10^5$--$10^{10}M_{\odot}$.} For example, in the center of our Galaxy
there is a black hole of $4 \times 10^6M_{\odot}$, and in the center
of the Andromeda Nebula---a hole with a mass of about
$10^8M_{\odot}$ \citep{Cherepashchuk2014, Bender2005}. The total number of SMBHs
in the Universe can be estimated from the number of large galaxies as $ 10^{11}$,
but their formation in the observed number and at the earliest stages
of the expansion of the Universe is an unsolved problem. Previously,
the generally accepted model was that dark matter at an early stage of the Universe
formed gravitationally bound clusters due to the Jeans instability, in which baryons
then accumulated \citep{Bullock2017}. Depending on the mass of the dark matter cluster,
these baryonic clusters produced globular clusters of stars, dwarf galaxies,
and massive galaxies---spiral or elliptical. After that, the galaxies formed
massive black holes from thousands to billions of times the mass of the Sun.
The presence of supermassive black holes in the early stages of the expansion
of the Universe, before the emergence of massive galaxies, indicates
an alternative scenario that has been gaining popularity in recent years.

In 2014, A.~M.~Cherepashchuk noted in a review on black holes: ``some scientists
seriously discuss the question of what is primary: the formation of a galaxy
in the early stages of the evolution of the Universe, followed by the formation
of a supermassive black hole in its center, or the formation of a primary supermassive
black hole, which then ``pulls'' the baryonic matter from which the stars of
the galaxy are formed'' \citep{Cherepashchuk2014}. For example, \citet{Carr2018} believe
that supermassive black holes can be the seeds for the formation of galaxies.
If the formation of galaxies began with accretion to the seed black hole,
then dark matter was attracted to the existing galaxy, increasing its mass.
This not only changes the generally accepted view of galaxy formation,
but also allows us to reconsider the formation of such large-scale cosmological
structures as galaxy clusters.

Thus, numerous observations prove that there is a vast population of black holes
in the Universe from $5M_{\odot}$ to $ 10^{10}M_{\odot}$.
Perhaps the observed population is responsible for the dark matter phenomenon
and for the formation of cosmic structures, including galaxies \citep{Clesse2017, Clesse2018, Garsia2018}.
These issues are discussed in detail in several recent reviews:
\cite{Cherepashchuk2016, Carr2018, Dolgov2018}. Note that the very fact of the existence
of black holes was proved by the image of the event horizon of the SMBH with the mass
\mbox{$6.5\times10^9M_{\odot}$} at the center of the galaxy M\,87 \citep{Akiyama2019}.
\citet{Carr2018} note the difficulty of developing a single mechanism
to create an observable distribution of black holes that differ so significantly
in mass and number. Numerous discoveries of small SBHs and supermassive SMBHs
are accompanied by unproductive searches for intermediate-mass black holes (IMBH).
Many authors believe that this reflects the two-peak distribution of black holes
associated with two different mechanisms of black hole formation and suggest that,
in addition to stellar evolution, there is another mechanism for the formation
of so-called primary black holes at an early stage of the Universe, for example,
from fluctuations in the density of hypothetical fields \citep{Dolgov2018}.
But so far, the evolution of massive stars, as well as the merger of two neutron stars,
are the only reliable scenarios for the emergence of black holes.

In recent years, there has been a growing interest in bounce cosmologies
and cyclic models of the Universe \citep[see][]{Brandenberger2017, Novello2008},
and it has been hypothesized that some of the black holes came from the past cycle
of the Universe, and the largest of them were the seeds for the formation of galaxies
\citep{Clifton2017}. Indeed, black holes are indestructible objects
that must survive both a single and multiple passage through the most compressed state
of the Universe. Conventional physics does not know the mechanisms of destruction
of black holes, the gravity of which prevents any of their disintegration,
regardless of any changes in the scale of the Universe. Apart from quantum Hawking evaporation,
the only process that destroys a black hole is its fusion with another hole,
which produces a larger hole and preserves the indestructibility of the entire black
hole population. Note that the quantum evaporation of a stellar-mass black hole
is possible only if the energy inflow to it is less than the outflow
in the form of Hawking radiation. Even without taking into account the radiation of stars,
such evaporation is possible only at an extremely low temperature of the relic radiation.
Since the term ``primary black holes'' is used in relation to black holes
that originated at the very beginning of a given cycle of the Universe,
we will use the term ``relict black holes'' (RBH) for black holes
left over from past cycles of the Universe. The paper by \citet{Clifton2017}
considers the transition of black holes through the maximally compressed state
of the Universe, which is characterized by a sufficiently low density.
As in our work, black holes that have passed from the past cycle of the Universe
to the current one appear as dark matter, and the most massive of them
are the seeds for the formation of galaxies.  The mathematical solution
obtained in the article \citet{Clifton2017} is illustrative and uses
a small \mbox{(5--640)} number of black holes, imposing on them the condition
of a sufficiently rare location to avoid their merging.
The model of the transition of black holes from the past cycle to the present one,
discussed in our article, differs from the \citet{Clifton2017} scenario
in that the merging of a large number of black holes in it is not only not prohibited,
but even necessary to trigger the mechanism of the Big Bounce or the Big Bang
(see Appendix). Note that numerous papers discussing the origin
of primordial black holes at the early stage of the Big Bang (see the review
\citet{Dolgov2018} and references therein), consider a variety of mechanisms
for the formation of such holes, but these works are fundamentally different
from those on relic black holes, which consider the transition of these holes
from the past cycle of the Universe, and not their formation during the Big Bang.
As a rule, the mechanisms of formation of primary black holes are considered
within the framework of a one-time model of the Universe, while the existence
of relict holes implies a cyclic cosmology, or at least a bounce cosmology.
Relict black holes, born in a certain cycle of the Universe, should,
during the transition from cycle to cycle, constantly decrease in number
due to mutual mergers, but grow in size, both due to mergers and due
to the accretion of the surrounding matter. Naturally, in each cycle,
a new population of black holes will be born, thus, in the cyclic Universe,
there must be a mixture of black holes of different ages---just as
the human population consists of people of different birth years.

To test the hypothesis of the existence of relic black holes, expressed by
\citet{Clifton2017}, we need to consider the following problem: are bounce cosmologies
and cyclic models of the Universe compatible with the presence
of a large observable population of black holes? Do numerous black holes impose
severe constraints on any model of an oscillating Universe? If black holes go
from cycle to cycle, how does their distribution change? Is it possible
to build at least the simplest model of the evolution of black holes
in an oscillating Universe?

This article is dedicated to discussing these issues. If the Universe
is an oscillating object with black holes of different ages,
then studying it should shed light on the following specific questions:
\begin{list}{$\bullet$}{
        \setlength\leftmargin{5mm} \setlength\topsep{2mm}
        \setlength\parsep{0mm} \setlength\itemsep{2mm} }
\item Where did such a large number of SBHs with a mass of up to 100 $M_{\odot}$ come from?
\item How did the very massive ($10^5$--$10^{10}M_{\odot}$) black holes appear,
which are observed even at the earliest stages of the Universe \citep {Banados2017}?
\item Why are there so few IMBHs with a mass of $10^2$--$10^5M_{\odot}$?
\end{list}
The purpose of this article is not to discuss the specific mechanisms
of the transition of the Universe from cycle to cycle. The literature describes
many scenarios for such a transition, both in Einstein's theory and in
non-Einstein models: \citet{Steinhardt2002,Novello2008,Penrose2011,Poplawski2016,
Gorkavyi2016,Brandenberger2017,Gorkavyi2018a,Gorkavyi2018b} (See also Appendix).
We are only interested in the evolution of the most indestructible
component of the Universe---black holes, considering the evolution
of their population within the framework of a simple iterative model.
For definiteness, as a basis for consideration, we will take the classical model
of the oscillating Universe, which was considered, for example, by the group
\citet{Dicke1965}. From this model it is possible to obtain an estimate
of the size of the periodic Universe, compressed to the radius
of photodisintegration of nuclei: about 10\,light-years.
Indeed, if we take the modern Universe with a size of the order of
100\,billion\,light-years and a temperature of the relict radiation
of about 3\,K and compress it by a factor of $10^{10}$ (to the size of
10\,light-years), then we obtain the radiation temperature of approximately
$ 10^{10}$\,K. Under such conditions, effective photodissociation
of the nuclei of heavy elements begins, which renews the chemical
composition of the Universe and makes it capable of forming stars in
a new cycle, as noted by \citet{Dicke1965}. We will consider
the evolution of the black hole population in this classical model
of an oscillating Universe, assuming that the findings are applicable
to many other cyclical models of the Universe.
  
\section{QUALITATIVE ESTIMATES OF THE TOTAL VOLUME OF BLACK HOLES}

If the oscillating Universe systematically passes through a squeezed state
with a size of about $ 10$ light-years, then how many black holes
can be placed in such a volume to ensure their complete or partial
transition from the old to the new cycle?

For the number of black holes, we assume values comparable
to observations and estimates of the mass of dark matter in the Universe.
The maximum number of SBHs with a mass of $5M_{\odot}$ is estimated
at $10^{23}$, which gives a total mass of $5\times 10^{56}$\,g.
The number of SMBHs can be estimated from the number of massive galaxies
as $10^{11}$. If we assume that the average mass of SMBH is $10^6M_{\odot}$,
then the mass of the SMBH population will be $10^{17}M_{\odot}$ or
$2\times 10^{-7}$ of the mass of SBHs.

Let's estimate in what minimum volume it is possible to place the population
of SBH and SMBH black holes, if we do not take into account their merger.
It is easy to show that the total volume of SMBHs is several orders
of magnitude larger than the volume of all SBHs. But even among SMBHs,
the contribution to the total volume is mainly given by holes of maximum mass,
not average. Without taking into account the merge, $10^9$ SMBHs with
a mass of $10^9M_{\odot}$ can be placed in a cube with an edge size
of about $ 1$ light-year, and $10^{11}$ SMBHs with a mass of
$10^6M_{\odot}$ will take $ 10^{7}$ times less volume.

Obviously, when the Universe collapses, black holes will merge.
If SBHs merge without gravitational radiation, then the resulting hole
will be comparable in size to the Universe itself; similar SMBH mergers
would yield the resulting black hole 5--6 orders of magnitude smaller,
but still huge. If gravitational radiation is taken into account,
then the estimate of the size of the final black hole will sharply decrease.
Merging two black holes with a size of $R$, taking into account the maximum
level of gravitational radiation, gives a final black hole with
a radius of $R_{\rm min}$ and an entropy equal to the sum of the entropy
(and surface) of the two initial black holes (see
\citet{Bekenstein1973,Hawking1975}): $R_{\rm min} ^2 = 2\times R^2$.
When merging $N$ black holes, we get
\begin{equation}
{R_{\rm min} ^2  =  N \times R^2 ~\text{or} ~   R_{\rm min}  =  R\times N^{1/2}}.
\end{equation}

Hence $R_{\rm min} \sim 0.6$\,light-years for $10^{23}$ SBHs with
a mass of $5M_{\odot}$, $R_{\rm min} \sim 1$~light-years for
$10^{11}$ SMBHs with an average mass of the order of $10^6M_{\odot}$ and
$R_{\rm min} \sim 10$\,light-years for $10^9$ SMBHs with a mass
of the order of $10^9M_{\odot}$. These estimates show that
if the minimum size of the Universe is close to ten light-years,
then the observed population of black holes can pass through
such a compressed state of the Universe and get into the next cycle.

\section{ITERATIVE MODEL FOR THE EVOLUTION OF BLACK HOLES}

Consider a conditional model of an oscillating Universe, described
by General Relativity or any other theory that has black holes.
The maximum size of the Universe is insignificant for us,
as is the period of the oscillations. We assume that the accelerated expansion
of the Universe and the specific mechanism of the transition from
its expansion to compression do not affect the solutions of
the simple model under consideration. Let's assume that the minimum size
of the compressed Universe is comparable to 10 light-years.
As the estimates given in the previous section show, this volume
is sufficient to accommodate all the black holes and move them
from one cycle to another. We believe that black holes occur only
during stellar evolution and during the merger of neutron stars.
Black holes increase their mass by accretion of surrounding matter
and radiation, and decrease their number by mutual fusion.
It is obvious that the stationary distribution of black holes,
the number of which increases as a result of the collapse of
ordinary stars and the merger of neutron stars, is achievable
only if there is some mechanism for destroying black holes or removing
them from the population. For example, \citet{Penrose2011},
in his cyclic model of the Universe, suggested that so much time passes
between cycles that Hawking radiation vaporizes even
the largest black holes. The Penrose model can be considered as
the limiting case in which the cyclicity is achieved by the complete
destruction of the black hole population. Trying to get
the most general results, we will consider the evolution of the distribution
of black holes in the most general form, describing
the phenomenological coefficient of the decrease in the number of black holes
of the same age. From a physical point of view, this coefficient
can describe not only the decrease in the number of black holes
during mutual mergers, but also any other mechanisms of black hole
disappearance. When discussing the resulting model, we will consider
what methods exist to achieve a stationary distribution of black holes.

Suppose that in the Universe in some cycle, which we will call zero,
as a result of stellar evolution, an initial population of identical
black holes with an individual hole mass of $m_0$ and a population of $n_0$
has emerged. A continuity equation with additional terms that are analogous
to ``chemical reactions'' \citep{Gorkavyi1994} can describe changes
in the concentration and mass of any population of objects,
including black holes with a given initial mass and initial concentration
\begin{equation}
\dfrac{\partial nm}{\partial t} + \nabla (n \times m \times V) = A (n,m) - B (n,m),
\end{equation}
where $A (n,m)$ is a term describing the increase in the concentration and mass
of black holes, and $B (n,m)$ is a term describing the decrease in these quantities.
This equation is difficult to solve, even if we neglect the terms of the equation
that depend on the speed. Consider the case when the change in the average mass
$m$ and the concentration $n$ are not related to each other.
For example, the mass of a black hole can change due to the accretion
of surrounding scattered matter or radiation. These processes are not related
to the concentration of black holes. For the case of independent processes
describing a decrease in the concentration of $n$ and an increase
in the mass of $m$, we can write:
\begin{equation}
\dfrac{\partial n}{\partial t} = -\dfrac{B(n,m)}{m},
\end{equation}
\begin{equation}
\dfrac{\partial  m}{\partial t} = \dfrac{A(n,m)}{n}.
\end{equation}
If we go to iterative formulas, then from equation (3) we get:
\begin{equation}
{n_{i+1}} - {n_i} = -D n_i,
\end{equation}
where the parameter $D$ should not depend on the mass:
\begin{equation}
D = \dfrac{B(n,m)}{nm} T.
\end{equation}

The mass of individual black holes over a given time interval increases
by an amount proportional to the surface area of the black hole, that is,
its mass squared. From equation (4) we write
\begin{equation}
{m_{i+1}} - {m_i} = C m_i^2,
\end{equation}
where the parameter $C$ should not depend on the concentration:
\begin{equation}
C = \dfrac{A(n,m)}{nm^2} T.
\end{equation}

For further analysis, we will use simple iterative laws of black hole
population evolution. Let the number of the initial population of black holes
born in the conditional zero cycle fall with each iteration of $i$ as

\begin{equation}
n_{i+1} = n_i(1-D),
\end{equation}
where $n_i$ is the number of black holes in the iteration $i$, D is
the coefficient that takes into account the decrease in the number of black holes
in each cycle. Equation (7) for the average individual mass is written
in the following form:
\begin{equation}
            m_{i+1}  =  m_i(1+ Cm_i),
                \end{equation}
where $C  =  (108\pi G^2/c^3)T\rho$, $T$ is the time interval
of the black hole feeding or absorbing the environment with a density of $\rho$.
Here we assume that the velocity of the medium is equal to the speed of light,
that is, the hole is fed by the surrounding radiation. If we consider the mechanism
of the Bondi--Hoyle--Lyttleton accretion \citep{Bondi1952}, then the expression for $C$
will be similar, with the replacement of the speed of light by the speed of sound
and with a change in the numerical coefficient. Since we will consider $C$
as a numerical parameter, the details of the physical mechanism that leads
to this value of $C$ are not essential for our model.
The law (10) at $C m_i \ll 1$ leads to the mass of the black hole
slowly changing during iterations (and cosmological cycles).
At $C m_i > 1$, the growth rate of the black hole will, on the contrary,
be very high. We assume that the growth conditions for all holes are the same,
and only the difference in the size of the holes determines the different rate
of their growth. The evolution of all population components of different ages
is considered as independent.

Thus, our simple iterative model is controlled by only two parameters:
the parameter $C$ controls the change in the individual mass of the hole,
and the parameter $D$---the change in the number of holes. The values
$D$ and $C$ can be interpreted in different physical models, but
from a mathematical point of view, these are parameters of the simplest laws
of black hole evolution, which can serve as a useful starting point
for the development of more complex models.

If the iteration number $i$ implies the number of the cosmological cycle,
then expression (10) begins to underestimate the growth of the black hole
at the nonlinear stage, when $C m_i >1$. Therefore, for the nonlinear stage,
a much shorter time interval is used for the iteration, so that $C m_i << 1$,
where $i$ will be the iteration number and not the cosmological cycle number.
To achieve this condition, the number of steps on the last loop
is assumed to be $10^8$.

We are monitoring a population that originated in the zero cycle and moves
in time (in cycles), but in each cycle the stars will generate new initial populations,
so we will be dealing with a continuous stream of black holes from cycle to cycle.
At any given time, for example, in the present cycle of the Universe,
we must observe around us a mixture of black holes from different cycles.
Similarly, demography studies how the population of people born
in a particular year changes over time. For a country with a constant number
and structure of population, a simple summation of the data for this population
can calculate the total population and its age distribution at the current moment.
The stationary distribution of black holes is ensured by their birth
during stellar evolution and by the mandatory mechanism of removal
of at least a part of the black hole population.

\section{RESULTS OF CALCULATING THE EVOLUTION OF BLACK HOLES}

Consider Model~1 with the parameters: \mbox{$n_0=10^{21}$;}  $m_0=5M_{\odot}$;
$C= 5\times10^{-4} M_{\odot}^{-1}$;  $D=0.05$
The accepted initial number of black holes is greater than that
given by modern theories of stellar evolution, but this number refers
to black holes that were born not to the present time, but for the entire cycle
of the Universe---up to the stage of maximum compression.
In addition, all the calculation results change linearly from $n_0$,
so we can choose any other initial value of $n_0$ and easily
recalculate the data. Therefore, we will keep this value $n_0$
for all the models under consideration. The value of $C$ accepted
in Model~1 corresponds to the value of the product $T\rho=0.6$,
where the feeding time $T$ is in years, and $\rho$ is in g\,cm$^{-3}$.
The accepted value of $D$ means that in each cycle the current population
of black holes loses 5\% of its population. The results of the calculations
are shown in Figure~1.

\begin{figure}
    \includegraphics[width=\columnwidth]{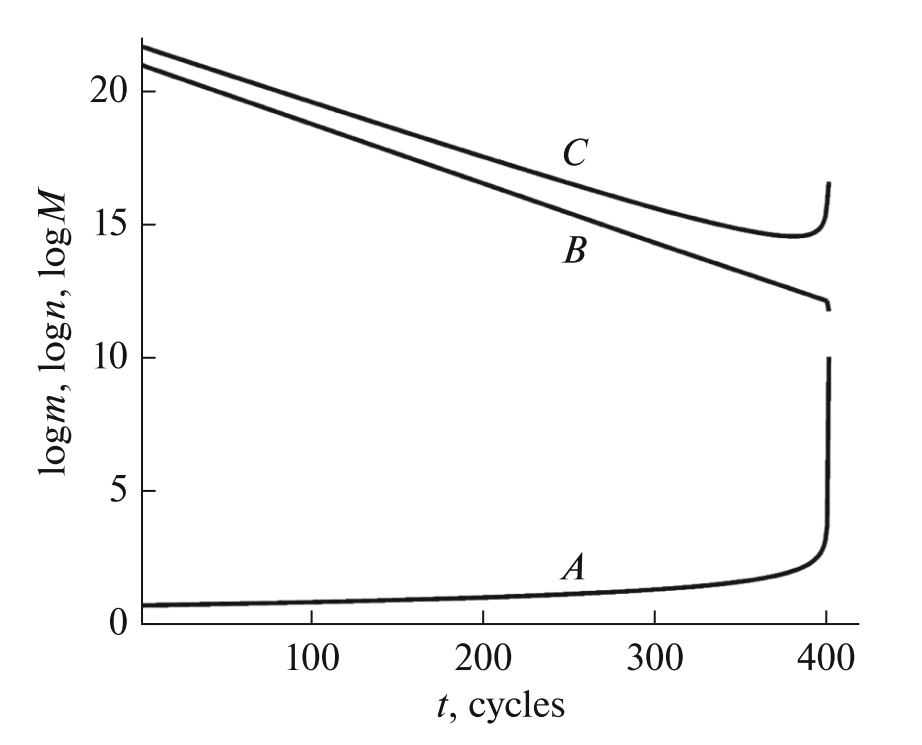}
    \caption{Evolution of the population of black holes in Model 1 over 402 cycles.
The individual black hole mass $m_i$ is shown by the lower curve A,
their number $n_i$ in each cycle---by line B. The total mass $M_i$
of the black hole population for each cycle is shown in graph C.
All masses are indicated in solar masses. A characteristic feature is
the rapid growth of individual weight in the last cycle.
The sharp decrease in the number of black holes in the last cycle
is due to the fact that it is not completed.}
    \label{fig:fg1} 
\end{figure}

\begin{table*}
\caption{Parameters of black hole populations in different models}
\label{Tab1}
\medskip
\begin{tabular}{l|c|c|c|c}
\hline
$Parameters$ & Model 1 & Model 2 & Model 3 & Model 4\\
\hline
                    \hline
~1~---$D$                             &0.05                &0.05               &0.04               &0.02              \\
~2---$C$, $M_{\odot}^{-1}$          &~~~$5\times10^{-4}$  &~~~$4\times10^{-4}$  &~~~$4\times10^{-4}$  &~~~$2\times10^{-4}$  \\
~3---$t_{\rm max}$, cycles           &401.41             &501.63             &501.63             &1002.33           \\
~4---Cycles for SBH                  &~~~~~~~~~~~0--381.40&~~~~~~~~~~~0--476.62&~~~~~~~~~~~0--476.62&~~~~~~~~~~~0--952.31         \\
~5---Cycles for IMBH                 &~381.41--401.39~   &~476.63--501.61~   &~476.63--501.61~   &~~952.32--1002.28~ \\
~6---Cycles for SMBH                 &~401.39--401.41~   &~501.61--501.63~   &~501.61--501.63~   &~1002.28--1002.33~ \\
~7---$m_{\rm av}$ SBH, $M_{\odot}$  &5.28               &5.22               &5.28               &5.28              \\
~8---$m_{\rm av}$ IMBH,$M_{\odot}$  &481                &440                &482                &479               \\
~9---$m_{\rm av}$ SMBH, $M_{\odot}$ &$1.158\times10^{6}$~~~~~~&$1.155\times10^{6}$~~~~~~&$1.155\times10^{6}$~~~~~~&$1.152\times10^{6}$~~~~~~\\
10---$m_{i-1}$, $M_{\odot}$        &4815               &3934               &3934               &15058              \\
11---$N_{\rm SBH}$                 &$1.9\times10^{22}$ &$1.9\times10^{22}$ &$2.4\times10^{22}$ &$4.9\times10^{22}$ \\
12---$N_{\rm IMBH}$                &$4.5\times10^{13}$ &$3.8\times10^{11}$ &$6.0\times10^{13}$ &$1.4\times10^{14}$ \\
13---$N_{\rm SMBH}$                &$2.6\times10^{10}$ &$2.0\times10^{8}$  &$3.6\times10^{10}$ &$8.2\times10^{10}$ \\
14---$M_{\rm SBH}$,  $M_{\odot}$   &$1.0\times10^{23}$ &$9.9\times10^{22}$ &$1.3\times10^{23}$ &$2.6\times10^{23}$ \\
15---$M_{\rm IMBH}$,  $M_{\odot}$  &$2.1\times10^{16}$ &$1.7\times10^{14}$ &$2.9\times10^{16}$ &$6.7\times10^{16}$ \\
16---$M_{\rm SMBH}$,  $M_{\odot}$  &$3.1\times10^{16}$ &$2.3\times10^{14}$ &$4.1\times10^{16}$ &$9.5\times10^{16}$ \\
17---$M_{\rm SMBH}/M_{\rm SBH}$    &$3.1\times10^{-7}$ &$2.3\times10^{-7}$ &$3.2\times10^{-7}$ &$3.7\times10^{-7}$ \\
18---$M_{\rm IMBH}/M_{\rm SMBH}$   &0.70               &0.74               &0.70               &0.71               \\
19---$R_{\rm BOX}$, light-years       &0.20               &0.04               &0.22               &0.28               \\
20---$R_{\rm SBH}$, light-years       &0.23               &0.23               &0.26               &0.37               \\
21---$R_{\rm SMBH}$, light-years      &1.7~~~             &0.1~~~             &1.9~~~             &2.9~~~             \\
\hline
\end{tabular}
\end{table*}

Fig.~1 shows a uniform decrease in the number of black holes with an increase
in the number of cycles of the Universe (line B) and an uneven increase
in the individual mass of black holes (curve A). The mass of the population
(curve C) falls during 380 cycles, but after the individual mass of black holes
(curve A) grows to $100M_{\odot}$, the total mass of black holes in this cycle
begins to grow. As follows from Fig.~1 and Table~\ref{Tab1}, the population
of black holes in Model~1 ends its evolution on the 402d cycle, leading to black holes
that exceed our accepted minimum size of the Universe in 10 light-years.
We will limit our consideration to black holes of no more than $10^{10}M_{\odot}$.
The model results show (see Fig.~1 and Table~\ref{Tab1}) that black holes
gain weight very slowly at first, accelerating rapidly at the end of their evolution.
To increase the initial weight of the SBHs 20 times---from $5M_{\odot}$ to
$100M_{\odot}$---it took 381 cycles, while further increasing the hole mass
by a factor of 1000---from 100 to $10^5M_{\odot}$ and the formation of IMBHs,
it took only 20~cycles. The formation of a population of supermassive black holes,
growing up to $10^{10}M_{\odot}$, occurs during $2\%$ of the duration of the last cycle.
Any accretion models for the growth of supermassive black holes
from black holes of intermediate masses described in the literature are included
in the model under consideration through the growth parameter $C$.
Our model confirms that the largest black holes are particularly efficient
at absorbing the environment.

Note that the time of the iterative model and the time of the real Universe
do not have a direct connection within a separate cycle. The time in the model
is based on the assumption that the density of the medium that feeds black holes
is constant. In fact, the density of this medium can change by many orders of magnitude
over the course of a cycle, so in reality $2\%$ of the model time may turn out
to be half the time of the cycle of the Universe occurring in low-density conditions.
But even models based on the average density of the feed medium per cycle
are very informative. Table~\ref{Tab1} shows the data for the four calculated models.
In the first column, the numbers indicate the parameters and the corresponding variables.
Parameter\,3 is the maximum time of evolution of this model, parameters\,4--6 are
the times during which black holes of different masses were formed.
The average mass of SBH (parameter\,7 in Table~\ref{Tab1}) in the first model has only
$5.3M_{\odot}$, and IMBH---$481M_{\odot}$ (parameter\,8), and SMBH---more than
a million solar masses (parameter\,9). The individual mass of black holes at the end
of cycle 401, which preceded the final cycle of the model, is $4815M_{\odot}$ ($m_{i-1}$,
parameter\,10), that is, until the very last cycle, only SBHs and IMBHs were formed.
In the real Universe, there will be a mixture of populations of different cycles.
Parameters\,11--13 reflect the total number of black holes of different classes,
and parameters\,14--16---the total mass of these classes. The total mass of SBHs
certainly dominates the other classes and is $10^{23}M_{\odot}$, which exceeds
the mass of the initial population \mbox{$5\times 10^{21}M_{\odot}$} by 20~times.
The mass of SBHs in model 1 is comparable to the mass of the Universe.

Thus, the cyclic Universe model offers a natural mechanism for the accumulation
of black holes, based on the relatively small number of black holes that occur
in each cycle as a result of the evolution of massive stars and the merger
of neutron stars. The number of SMBHs formed in this accumulation process
(parameter\,13) approaches the observed number of large galaxies.
The ratio of the mass of SMBH to the mass of SBH (parameter\,17) is
$3\times 10^{-7}$, which coincides well with the estimates obtained from
observations (see Section~2). The individual mass of IMBH is more than
three orders of magnitude lower than that of SMBH. Although there should be
at least 1000 IMBH per SMBH in each galaxy, the small individual masses of IMBH
make them difficult to detect. As parameter 18 shows, IMBH is also inferior to SMBH
in total mass. The search for lighter SBHs is significantly simplified
by the fact that they can be part of binary systems. Therefore, they are often surrounded
by a brightly luminous accretion disk. SBHs also often merge with each other,
generating a recorded burst of gravitational radiation.

Parameter\,19 is the radius of the ball, which can accommodate all the black holes
of the Universe obtained in Model 1 (without taking into account their merger).
The main volume there will be occupied by SMBHs. Parameter\,20 is the size
of the black hole that is obtained from all SBHs of the Universe,
with the maximum efficiency of gravitational radiation (that is, a hole
whose entropy or area is equal to the total entropy or area of all SBHs).
Parameter\,21 is the similarly obtained total black hole for SMBHs.

Model 2 differs from Model 1 by reducing the parameter $C$, that is,
the growth rate of black holes. This increases the duration of evolution
to 502\,cycles and compensates for the decrease in the growth of black holes
by increasing the accumulation time. As a result, the total amount and mass
of SBHs almost did not change, but the amount and mass of IMBHs and SMBHs
decreased by two orders of magnitude. This model does not correspond to reality,
but demonstrates the dependence of the results on the parameter $C$.

To return the SMBH population to more realistic values, in Model 3 we keep
$C$ from Model 2, but reduce $D$ to 0.04. This does not change the number of cycles
in the model, but has a positive effect on the populations of IMBHs and SMBHs,
which become even larger than in Model 1. The mass of SBHs also increases slightly.
In Model 4, we reduce $C$ and $D$ by half, compared to Model 3.
The number of cycles of the Universe increases to 1003, and the mass
of all populations increases by two or more times.

We assumed that the initial population consists of black holes of the same mass.
In reality, the evolution of stars of different masses will give a whole range
of sizes of black holes. However, these calculations indicate that the distribution
of black holes, which determine the bulk of the mass of the Universe, should be close
to the peak in the mass distribution in the initial population.

As follows from these calculations, any mechanisms for reducing the number
of black holes that are embedded in the coefficient $D$ must be accompanied
by a mechanism for removing the largest black hole that is formed during mergers.
As already mentioned, in the Penrose model, Hawking evaporation is responsible
for destroying the entire population of black holes, including the largest hole
\citep{Penrose2011}. In the models of the Universe, which is located in a huge black hole
\citep{Poplawski2016,Patria1972,Stuckey1994}, there is an interesting possibility
of getting rid of the emerging largest black hole: it can grow so much that
it will merge with the outer hole \citep{Gorkavyi2018a, Gorkavyi2018b}.

The size difference between SBHs and the entire Universe is 24 orders of magnitude.
Note the interesting matches indicated by the \mbox{parameters\,19--21}
of Table~\ref{Tab1} for the most realistic models\,1, 3, 4 (excluding data
for the demo Model\,2):

\begin{list}{}{
\setlength\leftmargin{5mm} \setlength\topsep{2mm}
\setlength\parsep{0mm} \setlength\itemsep{2mm} }
\item 1. From the observed temperature of the relic radiation and the photodissociation
requirement of atomic nuclei, we obtain an estimate of the size of the compressed
Universe of approximately 10\,light-years \citep{Dicke1965}.

\item 2. The ball of the minimum volume in which all the observed black holes
of the Universe can be packed has a radius of approximately 0.2--0.3\, light-years
(parameter\,19). This means that the process of mass merging of black holes
should occur just when the collapsing Universe reaches the radius of
photodissociation of the nuclei.

\item 3. The merger of all SBHs at the maximum efficiency of gravitational radiation
leads to the formation of a single hole with a radius of 0.2--0.4\,light-years
(parameter\,20).

\item 4. The populations of SMBHs and SBHs are completely different: the number of SMBHs
is less than the number of SBHs by almost 12~orders of magnitude,
and the total mass of SMBHs is 7~orders of magnitude less than that of SBHs,
although the individual mass of SMBHs exceeds that of SBHs by five orders of magnitude.
Nevertheless, at the maximum efficiency of the gravitational radiation,
the SMBH merger leads to the formation of a final hole with a radius
of 2--3 light-years (parameter\,21), which is comparable to the size of the Universe,
at which the destruction of the nuclei of chemical elements by gamma quanta
of relic radiation begins.
\end{list}

The coincidence of several parameters so different in physics may indicate
that the Universe is a self-adjusting system that has evolved as a result
of many cycles to the most optimal parameters. It is logical to assume
that during the compression of the Universe, the most effective mechanism
for turning black holes into gravitational waves is realized.
Indeed, let the black holes merge with less efficiency and produce a final black hole
with a size much larger than 10 light-years. From the point of view
of a freely falling observer easily penetrating the interior of this large black hole,
nothing prevents the smaller black holes inside it from continuing
to generate gravitational waves up to the theoretical limit.
The high efficiency of the transformation of collapsing matter
into gravitational waves will be discussed in the article \citet{Gorkavyi2020}.

\section{OBSERVATIONAL CONFIRMATIONS OF THE ITERATIVE MODEL}

Let us compare the results of iterative models for the evolution of black holes
with the parameters of black holes of different classes: SBH, IMBH, and SMBH,
which should be the main observational test of the models under discussion.

\subsection{Dark Matter and the Number of Galaxies}

Models of the evolution of black holes in an oscillating Universe
(Table~\ref{Tab1}), despite their simplicity, consistently reproduce
the two main observed features of the black hole population:

\begin{list}{}{
\setlength\leftmargin{5mm} \setlength\topsep{2mm}
\setlength\parsep{0mm} \setlength\itemsep{2mm} }
\item 1. A large number of stellar mass black holes (SBHs) up to $100M_{\odot}$,
which accumulates over the course of many cycles. This corresponds to
LIGO observations and may explain the existence of dark matter---partially or completely. This eliminates the need for an additional mechanism
for the formation of black holes, other than the collapse of massive stars.

\item 2. The existence of a population of supermassive black holes of the order of
$10^{11}$, comparable to the number of galaxies and making up about $ 10^{-7}$
of the mass of all SBHs, or the entire mass of the Universe, which is close
to the observed values.
\end{list}

Many experts believe that the hypothesis that dark matter consists
entirely of black holes and neutron stars is in contradiction
with the observational data on the gravitational lensing of stars
in the Milky Way bulge and in the Magellanic clouds (see the review
and references in the articles \cite{Dolgov2018}, \cite{Belotsky2019}
and \cite{Carr2020}). Indeed, if black holes or neutron stars
are evenly distributed around a Galaxy, they should block out
the ordinary stars of neighboring galaxies, which will cause
an occasional increase in the luminosity of individual stars.
But if black holes tend to form clusters, then this observational constraint
becomes invalid \citep[see the analysis of ][]{Clesse2017, Clesse2018, Belotsky2019}.
According to the review \citep{Carr2020},
the various observational constraints for black holes in the region
of 4--10$M_{\odot}$ are minimal. We believe that dark matter may consist entirely
of holes of this mass. Note that each observational constraint
is accompanied by a series of theoretical assumptions that require
careful analysis, so it is impossible to perceive these constraints
as reliably established. Is it possible to reconcile the frequent
black hole mergers observed by LIGO with the small number
of gravitational lensing events? We have done
an independent estimate of the frequency of black hole mergers was made
within the framework of the Clesse and Garcia-Bellido hypothesis.
It is shown that the LIGO data on the frequency of black hole mergers
are well explained by the assumption that a galactic dark halo
with a mass of $10^{12}M_{\odot}$ consists of $2\times10^{11}$ black holes
with masses of $5M_{\odot}$, if this population of black holes
is collected approximately in $10^6$--$10^8$ dark globular clusters
with masses of the order of $10^4$--$10^6M_{\odot}$. The term
``dark globular clusters'' (or ``dark star clusters'') was introduced
in \cite{Taylor2015}, where a class of globular clusters with
an abnormally high ratio of~$M/L$ was discovered near the galaxy Centaurus\,A.
\citet{Taylor2015} believe that these dark clusters may contain
large amounts of dark matter or an intermediate-mass black hole.
We believe that the dark clusters found near the galaxy Centaurus\,A
represent only a small visible part of the vast population
of inconspicuous black hole clusters that make up the dark halo of galaxies.
If the radius of a dark globular cluster is approximately 10\,light-years,
and they are located at a distance of about $10^{5}$\,light-years
from the center of the Galaxy, then the total area of dark globular clusters
in the sky will be approximately \mbox{65--6500\,$\Box\degr\!.$}
The uniform distribution of dark globular clusters throughout
the entire celestial sphere with an area of $41253\, \Box \degr$
means a small (0.1--10\%) probability that such a cluster
will fall on the line between the telescope and, for example,
the stars of the Magellanic clouds. Therefore, observations
of gravitational lensing in local areas of the sky will not be able
to register clusters of black holes, although these observations
can provide an estimate of background, scattered black holes
that did not enter the clusters (or flew out of the clusters
after mutual merger). Note that such dark globular clusters can cause
the effect of lensing both by the total gravitational field and
by the field of individual holes that are part of them.
Studying the motion of stars from astrometric catalogs (and Gaia data)
will help find dark globular clusters in the disk of the Galaxy.

If the accumulation of SBHs helps to solve the problem of dark matter,
then the second point, according to which a significant amount of SMBHs
existed immediately after the Big Bang, indicates in favor of a new paradigm
of the formation of galactic structures and the galaxies themselves,
which grew around supermassive black holes \citep{Cherepashchuk2014}.
In the traditional picture (galaxies are formed before SMBHs),
there are serious problems: supermassive black holes in the centers of galaxies
were discovered at such an early stage of the Universe
when their formation is problematic.

The density perturbations of a dark matter medium is strongly dependent
on the wavelength. Therefore, for every Milky Way-type galaxy,
there must have been about 500 dwarf satellite galaxies (see \citet{Klypin1999}).
But only a few dozen satellites have been found near our Galaxy.

Ancient globular clusters do not move in the disk, but in the spherical halo
of the galaxy \citep{Payne1979}. The number of globular clusters
that grew out of relatively small fluctuations of dark matter
or some hypothetical field must be very large. For example,
\citet{Dolgov2018} examines the formation of primordial black holes
from fluctuations in the gravitating medium and concludes
that for every SMBH in the center of the galaxy, there must be
$10^5$--$10^6$ intermediate-mass black holes. This number is
several orders of magnitude greater than the number of globular clusters
in our Galaxy, of which there are less than 200. The record
for the number of globular clusters belongs to one of the largest
elliptical galaxies NGC\,4874---$3\times 10^4$ clusters,
but that's all is much less than the value predicted by \citet{Dolgov2018}.

The existence of relict SMBHs explains the large number of quasars
and massive holes in the centers of galaxies in the earliest stages
of the expansion of the Universe. In addition, relict black holes
and their associations may be related with large-scale inhomogeneities
of the relict radiation, for example, with an anomalously cold spot
with a size of about $ 10\degr$ in the Eridanus constellation
\citep{Vielva2010, Akrami2019}.

The new paradigm of galaxy creation changes the view on the deficit
of galactic satellites and the formation of globular clusters
\citep{Cherepashchuk2014}. According to the new approach, the number
of bright globular star clusters, galaxies and their satellites
depends not on fluctuations in dark matter, but on the distribution
of intermediate-mass black holes. Let us compare the amount of IMBHs
obtained as a result of modeling (Table~\ref{Tab1}) with the observed number
of satellites of galaxies and globular clusters in the Milky Way.

\subsection{Satellites of Galaxies and Globular Clusters}

In the considered simple models, the total mass of IMBHs is about
0.7~mass of SMBHs, and the amount of IMBHs is $1700$ times greater
than the amount of SMBHs (see Table~\ref{Tab1}). We do not include
in these estimates the demo Model 2 with an underestimated number
of massive black holes. At the same time, the individual masses of IMBHs
are three orders of magnitude less than that of SMBHs, which makes
their detection problematic. Note that there are 157 globular clusters
in the Milky Way with a mass of $10^4$--$10^6M_{\odot}$. Relict black holes
with a mass of $10^3$--$10^4M_{\odot}$, which came from past cycles
of the Universe, are ideal candidates for the role of centers of formation
of such globular clusters. At the same time, this makes it difficult
to detect IMBHs, because the centers of many globular clusters
are too bright to detect a black hole there.

Let us calculate for the most realistic Model~4 the number of black holes
in nine intervals (in $M_{\odot}$) [5;\,50], [50;\,500], ... ,
[$5\times 10^8$;\,$5\times 10^9$] (see Fig.~3 and Table~\ref{Tab2}).

\begin{figure}
    \includegraphics[width=\columnwidth]{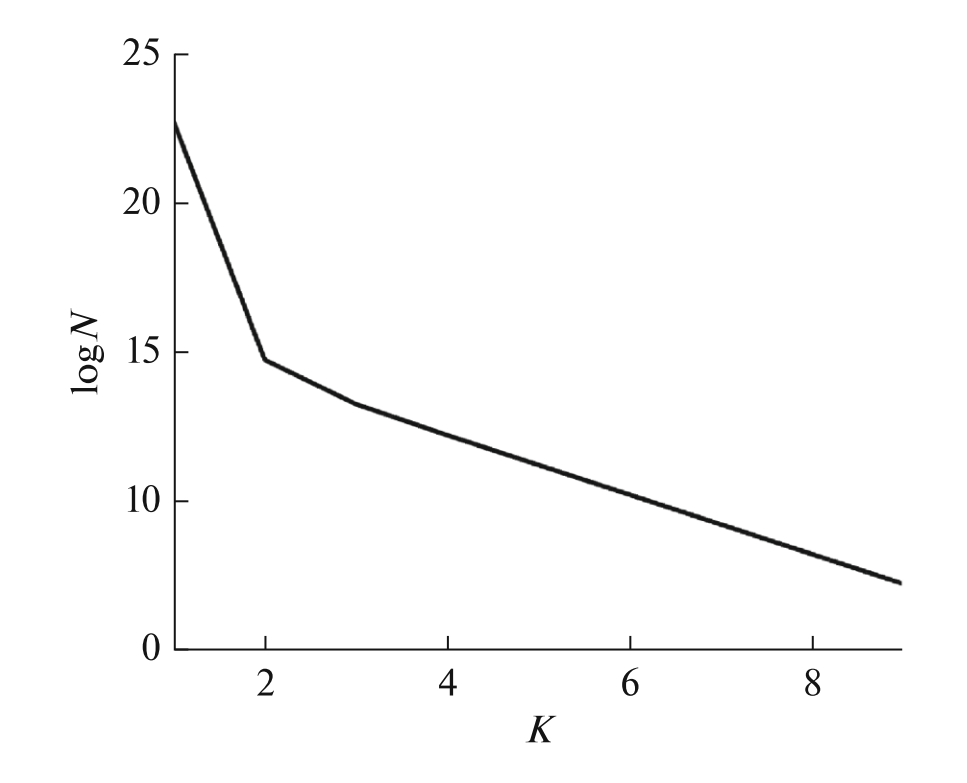}
    \caption{Black hole mass distribution in Model 4 over nine intervals
with the number $K$ (see Table~\ref{Tab2}).}
    \label{fig:fg1} 
\end{figure}

\begin{table*}
\caption{The number of black holes of various masses (Model 4)}
\label{Tab2}
\medskip
\begin{tabular}{c|c|c|c}
\hline
$K$ & Mass interval, $M_{\odot}$ & The number of black holes & Total mass, $M_{\odot}$\\
\hline
1 & ~5--50 & $ 4.9 \times 10^{22}$ & $1.6 \times 10^{23}$ \\
2 & ~~50--500 & $5.1 \times 10^{14}$ & $3.8 \times 10^{16}$ \\
3 & ~~~~~~~$500$--$5 \times 10^{3}$ & $1.7 \times 10^{13}$ & $1.0 \times 10^{16}$ \\
4 & $5 \times 10^{3}$--$5 \times 10^{4}$ & $1.5 \times 10^{12}$ & $0.9 \times 10^{16}$ \\
5 & $5 \times 10^{4}$--$5 \times 10^{5}$ & $1.5 \times 10^{11}$ & $0.9 \times 10^{16}$ \\
6 & $5 \times 10^{5}$--$5 \times 10^{6}$ & $1.5 \times 10^{10}$ & $0.9 \times 10^{16}$\\
7 &  $5 \times 10^{6}$--$5 \times 10^{7}$ & $1.5 \times 10^{9}$~~ & $0.9 \times 10^{16}$\\
8 &  $5 \times 10^{7}$--$ 5 \times 10^{8}$ & $1.5 \times 10^{8}$~~ & $0.9 \times 10^{16}$\\
9 &  $5 \times 10^{8}$--$5 \times 10^{9}$ & $1.5 \times 10^{7}$~~ & $0.9 \times 10^{16}$\\
\hline
\end{tabular}
\end{table*}

As can be seen from Fig.~2 and Table~\ref{Tab2}, the lion's share of black holes
(both in number and in total mass) falls on the interval 5--50$M_{\odot}$.
In the next mass interval, black holes are eight orders of magnitude smaller.
The dependence of the number of black holes with masses over $500M_{\odot}$
on the mass interval becomes linear. This weak dependence of the number
of IMBHs and SMBHs on their mass makes it possible to solve the problem
of missing satellites of galaxies and a small number of globular clusters.
Let us divide the population of black holes into four classes with larger mass
intervals based on the summation of the intervals from Table~\ref{Tab2}:
black holes that make up dark matter; holes in the centers of globular clusters;
holes in the centers of dwarf galaxies---satellites of a large galaxy;
SMBHs in large galaxies (Table~\ref{Tab3}). For clarity, we normalize
the number of black holes in these classes to the number of SMBHs
in large galaxies.

\begin{table*}
\caption{Relative abundance of black holes of different classes}
\label{Tab3}
\medskip
\begin{tabular}{c|c|l}
\hline
Mass interval, $M_{\odot}$ & Number of black holes & Observed objects\\
\hline
~~~~$5$--$500$ & $ 3 \times 10^{12}$ & Dark matter\\
~~~~~~~$500$--$5\times 10^{3}$ & $10^{3}$ & Globular clusters\\
$5 \times 10^{3}$--$5 \times 10^{5}$ & $10^{2}$ &Galactic satellites\\
$5 \times 10^{5}$--$5 \times 10^{9}$  & 1 & Large galaxies\\
\hline
\end{tabular}
\end{table*}

The number of globular clusters and dwarf satellites of large galaxies
per supermassive hole Table~\ref{Tab3}) is in good agreement
with observations. Clarification of the boundary between different classes
should also take into account the physical differences in the conditions
for the formation of globular clusters and satellites
of different types of galaxies. Note that, on the basis of observations,
an assumption was made about the presence of an intermediate-mass black hole
of the order of $3\times 10^3M_{\odot}$ in a globular cluster M\,15
and a black hole of $2\times 10^4$~$M_{\odot}$ in the Mayal\,II cluster
in Andromeda \citep{Marvel2002,Ma2007}. Globular clusters of a large number
of old stars that move not in the galactic disk, but in a spherical halo,
the main part of which apparently consists of stellar-mass black holes,
are logically explained by the formation of globular clusters
around intermediate-mass black holes (see the article \cite{Dolgov2017}
and references therein).

Globular clusters turn out to be just as old (or even older) than
the galaxies themselves and are attracted to them (and captured),
just like the SBH clusters in the galactic halo. This explains
why the mass of globular clusters is directly proportional to the mass
of the galaxy (which is close to the mass of its halo),
and the spatial distribution of old globular clusters coincides
with the halo of black holes. The proximity of the distribution
of globular clusters to the dark matter halo is another argument
in favor of the fact that dark matter consists of black holes.

\section{ARE THERE NEUTRON STARS OF PAST CYCLES?}

In classical cyclic cosmology \citep{Peebles1993}, it was assumed
that in the aggressive environment of a compressed Universe,
only elementary particles could survive, and the rest of the objects,
from the nuclei of atoms to stars, would be destroyed by photodissociation.
Obviously, black holes cannot be destroyed by photodissociation.
But what happens in a compressed Universe with neutron stars~(NS)?
Their density is comparable to the density of the atomic nucleus,
and they have such a large gravity that they can survive in the environment
of a compressed Universe. The question of the survival
of relict neutron stars (RNS) can be studied both from a theoretical
and observational point of view, but in this paper we will only talk
about possible observational manifestations for such objects.
Based on the total number of massive galaxies in the Universe
of the order of $10^{11}$ and the expected number of neutron stars
in a large galaxy of about $10^9$, we can roughly estimate
the maximum total volume occupied by all neutron stars in the Universe
as~ $10^{39}$\,cm$^3$ (linear size of the order of $10^{-5}$~light-years).
This volume, as well as the total number of neutron stars, is not comparable
to the number and total volume occupied by black holes.

Here are some arguments for the survival of neutron stars in a compressed
Universe. This survival depends primarily on the rate of photodissociation
and on the time when the Universe is compressed enough for photodissociation
to be possible. In the classical oscillating model, when the Universe
is compressed to a size of several light-years, a temperature
of the order of $ 10^{10}$K is reached, which is sufficient
for the photodissociation of the strongest nuclei---such as iron.
The maximum binding energy per nucleon in an iron core is about
$ 10$\,Mev. It is easy to show that the powerful gravity of a neutron star
leads to a gravitational binding energy of about $100$\,Mev per nucleon.
Thus, a neutron star is less susceptible to photodissociation
and can survive in the hot environment of a compressed Universe.
Note that the gravitational binding energy for white dwarfs is 2--3 orders
of magnitude less than that of neutron stars, thus white dwarfs
must dissolve in the environment of energetic gamma-rays.

Neutron stars may melt due to photodissociation on the Maxwell tail
of the gamma-ray distribution, and their mass may decrease.
For a single neutron star, the mass of the melted remnant,
which may still remain a neutron star, is considered, for example,
in \cite{Blinnikov1984,Landau1987}. It is shown that the minimum
possible mass of a neutron star can be approximately $0.1$--$0.2M_{\odot}$,
and its linear size will be several times larger than the size
of an ordinary neutron star. Thus, relict neutron stars,
which have a wide distribution of possible masses, can be left to us
as a legacy from the previous Universe. An important consequence
of the presence of neutron stars in the compressed Universe
is the active formation of black holes that occur when neutron stars merge.
This scenario was confirmed by the LIGO observations \citep{Abbot2017}.

\subsection{Number of Pulsars and Neutron Stars}
The number of pulsars in the Milky Way is of the order of\,$10^6$ \citep{Smith1979}.
\citet{Kippenhan1990} states: ``At present, such a number of pulsars
is already known that it is possible to assume the existence of about
a million active pulsars in our Galaxy alone. On the other hand,
observations of distant galaxies have been conducted over the past few decades
in order to determine how many supernova explosions occur on average
per century. This allows us to draw a conclusion about how many
neutron stars have appeared since ancient times in our Milky Way.
It turns out that the number of pulsars significantly exceeds
the number of neutron stars that could be formed as a result
of supernova explosions. Does this mean that pulsars can occur
in a different way? Perhaps some pulsars are formed not as a result
of star explosions, but in the course of less spectacular,
but more orderly and peaceful processes?'' In recent years, various solutions
to the problem of pulsar excess have been proposed,
but if the observed number of pulsars or neutron stars is indeed greater
than theoretically expected, then the hypothesis of possible preservation
of some neutron stars from the past cycle of the Universe
solves this problem without involving new mechanisms
for the formation of neutron stars.

\subsection{A Gap in the Mass Distribution of Neutron Stars and Black Holes}

NSs and BHs arise mainly at the end of the evolution of Wolf-Rayet stars
with masses of 5--50$M_{\odot}$ \citep{Ozel2016}. Therefore, it is expected
that the total mass distribution of neutron stars and black holes will form
a fairly smooth function, only with the difference that neutron stars
will have masses less than 3$M_{\odot}$, and black holes---more.
The exact value of the boundary mass between these two classes of space objects
is debated, but it is not essential for the smoothness of the expected
distribution function. However, observations show that neutron stars
have masses of 1--2$M_{\odot}$, and the masses of all black holes lie
in the region of more than 4--5$M_{\odot}$  \citep{Ozel2012}.
If black holes and neutron stars come from past cycles of the Universe,
then their distribution undergoes a significant change from the initial one:
black holes increase their mass, and neutron stars may decrease it.
This could explain the formation of a gap in the observed mass distribution
of black holes and neutron stars. Note that the simple models discussed
in the previous sections describe the growth of black holes
that depends only on their mass and the constant average density
of the feed medium. In reality, black holes and neutron stars fall
into conditions of high density and increased friction during
the collapse of the Universe, which can lead to mutual mergers, for example,
of neutron stars and an increase in the mass of the resulting star about twice,
which can lead to the formation of a black hole.
A similar mass jump in the first cycle can be experienced by merging
black holes that have arisen in star clusters. This effect can shift
the peak of the initial distribution of SBHs towards larger masses.
Our model does not describe these processes, because in them
the growth of body weight depends on their concentration.
Since the processes of jump-like growth as a result of mutual fusion
are essential only for bodies of small masses, they do not have
a significant effect on the entire spectrum of black holes.
Our iterative model implicitly takes this effect into account
by setting the minimum initial masses to 5$M_{\odot}$, which is more
than the minimum black hole mass of 3$M_{\odot}$ expected from the theory.

\subsection{Distribution of Pulsars and Neutron Stars by Periods and Velocities}
\cite{Narayan1990} published an analysis of the distribution of 300 pulsars
by rotation periods and velocities. They showed that pulsars are divided
into two populations: slowly (SR) and fast (FR) rotating objects.
The rotation period of fast pulsars is approximately equal to $ 0.2$~,
and of slow---more than $ 0.5$~s. The number of SR and FR pulsars:
$45\%$ and $55\%$. The speeds of movement of fast and slow pulsars turned out
to be interesting: 60 and 150 km\,$s^{-1}$. Thus, fast pulsars gravitate
towards the galactic disk, and slow pulsars---toward a spherical halo.

Although it is quite difficult to study the motions of pulsars in the Galaxy
due to the weak knowledge of their velocities, it is logical to assume
that it is in the distribution of pulsars and neutron stars by their velocities
and rotation periods that signs of the existence of relict neutron stars
from the past cycle of the Universe can be found. This raises the question
of the lifetime of a neutron star in the pulsar state and the existence
of mechanisms for maintaining them in this state. The same questions
about the mechanisms of radio emission were raised by researchers
in connection with the relatively recent discovery of rotating radio
transmitters (RRAT) \citep{McLaughlin2006}. There are attempts to explain
their sporadic pulsed radio emission, if they are old objects, for example,
by the matter falling on the pulsar from the supernova remnant \citep {Li2006},
by the reactivation of inactive vacuum gaps near the surface
of the neutron star by matter from the asteroids \citep{Cordes2008}.
There are other hypotheses, and their essence is that matter
falling on a neutron star can re-launch the neutron star as a radio pulsar.
Thus, despite the fact that the age of these relict pulsars is equal
to the age of the Universe, some of them may be active in the radio range.

\subsection{Single Relict Neutron Stars}

Neutron stars that have become pulsars before the next compression of the Universe
or during this compression, or have passed into the pulsar stage
due to the processes taking place in the compressed Universe, will evolve.
The evolutionary tracks of pulsars on the diagram ``period from the derivative
of the period'' ($P/\dot P$) (see, for example,
\cite{Phinney1981,Malov2001,Jonston2017} and references therein) indicate
that the period of the pulsar increases with time, and the magnetic field strength
(dipole component) decreases. The location of the pulsar on the $P/\dot P$ diagram
is approaching the so-called death line. The active lifetime of a radio pulsar
is estimated within a very wide range, but the typical lifetime is usually given as
\mbox{$10^6$--$10^7$\,years} (see \citet{Lorimer2005}). However,
as noted in Section~6.3, some single neutron stars can remain radio pulsars,
despite the age comparable to the age of the Universe. In any case,
these objects must lie behind the death line on the $P/\dot P$ diagram,
or be very close to it.

There are two main theories about the evolution of the angle between
the rotation axis and the magnetic axis of the pulsar. According to these theories,
in the course of evolution, these axes will coincide, and a coaxial rotator
will be observed, or the angle between these axes will be $90^{\circ}$
and an orthogonal rotator will be observed (see, for example,
\cite{Arzamasskiy2017} and references there). Even if there are factors
that will cause a new spin of the relict neutron star, turning it back
into a pulsar, this should not affect the established angle between the axes.
Hence, the RNSs will be either coaxial or orthogonal rotators.

According to theoretical studies, at the time of birth, a neutron star
has a high surface temperature ($10^{11}$\,K), but due to neutrino cooling,
this temperature drops rapidly and should reach about $10^8$\,K in a month.
The surface temperature of neutron stars in which it was estimated is
\mbox{$10^5$--$10^6$\,K} (see the review paper \cite{Yakovlev2004}).
The equation of state of a neutron star is poorly known, but it is obvious
that due to the lack of an additional energy source in the star,
the temperature will continue to fall over time. The age of relict
neutron stars exceeds the age of the Universe in this cycle,
and therefore their surface temperature should be lower than that
of ordinary neutron stars. Note that the final temperature of neutron stars
will also depend on such factors as the existence of long-lived
accretion disks around them.

\subsection{Relict Neutron Stars in Close Binary Systems}
If, for some reason, an RNS gets an accretion disk, or finds itself
in a close binary system with the transfer of matter from the companion
to its surface, it can become an X-ray pulsar. This pulsar will have
a specific set of properties that distinguish it from a regular pulsar.
As with single RNSs, the surface temperature of this pulsar will be lower
than that of ordinary pulsars. It will be either a coaxial or orthogonal
rotator. However, if its mass is small in comparison with the mass
of an ordinary pulsar, it will slow down very quickly, i.e. it will have
a large derivative of the period. The determined mass of such a pulsar
will be statistically less than it is for ordinary pulsars,
and the period of rotation in the system may be less
than for ordinary binary systems.

\subsection{Other Features of Relict Neutron Stars}

As is known, ordinary neutron stars are located in the flat component
of the Galaxy and are formed as a result of supernova explosions.
In the course of a supernova explosion, a shell with a mass significantly
greater than that of the resulting neutron star is shed, and, as a rule,
the remnant receives an additional impulse. As a result, the velocity
of the resulting neutron star can differ significantly from the velocity
of nearby stars. Therefore, a significant part of the pulsars
is distributed in a disk that is about 3 times thicker than
the disk of stars \citep{Payne1979,Arzamasskiy2017}.

Unlike ordinary NSs, RNSs should belong to the spherical component
of the Galaxy, and their velocity should not generally differ
from the velocities of neighboring stars and SBHs. Moreover,
some of these relict stars may not be part of the galaxies and will remain
in intergalactic space.

If RNSs of past cycles exist, then they should create a certain excess
of NSs, which should be especially noticeable at large galactic latitudes.

It is known that the number of binary systems in the Galaxy is comparable
to the number of single stars. RNSs, like ordinary stars, can be part
of binary systems. However, the pair ``RNS\,+\,ordinary star'' may have
features in comparison with the pair ``ordinary star\,+\,ordinary star''.
The mass of an RNS is comparable to that of a Sun-type star,
and the luminosity is very small due to the small surface area.
Therefore, an RNS can be searched for in such binary systems
where one companion is visible and represents the usual stellar population,
and the second companion has a mass of the order of
\mbox{$0.1$--$0.2M_{\odot}$} and is invisible in the optical range.
Mass detection of such systems will statistically support the hypothesis
of the existence of RNSs. Different scenarios of the formation
of a binary system involving a relict neutron star (or a relict black hole)
are possible, but we believe that the most likely scenario is that
an RNS captures so much matter from the gas-dust cloud
that a massive disk forms around it. The outer part of this disk
turns into an ordinary star, and the long-lived inner part of
the accretion disk feeds the relict neutron star, turning it into a pulsar.

\subsection{Conclusions on the Relict Neutron Star Hypothesis}

Let's summarize the contents of Section 6. Single RNSs will be located
in a spherical halo and have a velocity close to the velocity
of nearby stars and SBHs. The surface temperature and magnetic field strength
of the RNSs will generally be lower than that of ordinary neutron stars,
and the rotation periodswill be longer. Since most of these stars
are not pulsars, it is impossible to detect RNSs by classical methods
by searching for their periodic radiation. The most obvious way
of searching is to search for their blackbody radiation, for example,
in data on compact X-ray sources with constant flux density.
Due to the small surface of RNSs, their luminosity is low, therefore,
instruments with a very high sensitivity are needed for detection.
High sensitivity is expected in new space missions: ``Spectr--RG'',
launched in 2019, will scan the sky in the X-ray range,
``Spectr--UV'' -- in the ultraviolet range \citep{Boyarchuk1993},
and ``Spectr--M'' will observe in millimeter range with unprecedented
sensitivity \citep {Kardashev2000}. The RNS blackbody radiation,
if it can be detected, should be characterized by the fact
that its maximum will most likely be in the optical range.

If the RNSs are in binary systems, and the distances are such that
there is no accretion, then the search for RNSs should be carried out
on the visible companion, which will give an estimate of the mass
of the invisible star. Such systems may be good candidates
for inclusion in a single RNS search program. It is possible
that the data obtained in the optical range for about a billion stars
in the Gaia project will be a good basis for such a search.

If RNSs are part of close binary systems and accretion occurs
on the surface of a neutron star, then there is a chance that
the neutron star will turn into an X-ray pulsar. The mass of such a pulsar
can be abnormally small. It will be a coaxial or orthogonal rotator.

Relict neutron stars, both single and in binary systems, are best searched for
at high galactic latitudes, where the probability of detecting
an ordinary neutron star or pulsar is much lower than detecting
them in the Galactic plane. The discovery of a statistically significant
excess of neutron stars at high galactic latitudes
will provide additional indirect evidence in favor of the hypothesis
of the existence of RNSs.

\section{CONCLUSIONS}

A number of cyclic models of the Universe are discussed in the literature.
Almost none of them touches on the topic of the evolution of the population
of black holes from cycle to cycle. In this paper, we draw attention
to this problem and demonstrate that even simple models can shed light
on the occurrence of the observed mass spectrum of black holes.
Stellar evolution generates the initial distribution of black holes,
which, as they evolve and accumulate in the oscillating Universe,
multiply their number and mass. The simple models considered in this paper
qualitatively explain the origin and the ratio of the number and mass
of the three main populations of black holes: SBH, IMBH, SMBH.
\begin{list}{}{
\setlength\leftmargin{5mm} \setlength\topsep{2mm}
\setlength\parsep{0mm} \setlength\itemsep{2mm} }
\item 1. SBHs (less than $10^2M_{\odot}$), accumulating in the course of numerous
cycles, form a population that explains the LIGO observations and is probably
responsible for the dark matter phenomenon -- partially or completely.

\item 2. The mass of SMBHs ($10^5$--$10^{10}M_{\odot}$) in the considered models
is $(3$--$4)\!\times\!10^{-7}$ of the mass of SBHs. SMBHs may have existed
at the earliest stages of the expansion of the Universe and may have made
an important contribution to the formation of quasars and galaxies
that may form around SMBHs at the earliest stages \citep{Tang2019}.
This brings new light to the problem of reionization of the early Universe
\citep{Barkana2001}.

\item 3. The population of IMBHs ($10^2$--$10^5M_{\odot}$) lost its collective
number and total mass compared to SBHs during multiple cycles.
But IMBHs became the progenitor of the fast-growing SMBHs,
significantly inferior to them in terms of collective and individual masses.
Apparently, it is the IMBHs that are responsible for the formation
of globular clusters and the formation of satellite galaxies.

\item 4. The location of the ancient globular clusters associated with IMBHs,
not in a flat galactic disk, but in a spherical halo, confirms
the hypothesis that the galactic halo (that is, the dark matter halo)
consists of stellar-mass black holes.
\end{list}

The good agreement of the simple model of black hole evolution
with observations is a strong argument in favor of a cyclic model
of the Universe. Black holes in the oscillating Universe turn out to be
the determining factor for the formation of the hierarchy
of cosmic structures after the next Big Bang: from dark matter halos
from small holes of tens and hundreds of solar masses, to globular clusters
that grow around holes of thousands and tens of thousands of solar masses;
from dwarf galaxies with central black holes of the order of hundreds
of thousands of Solar masses to galaxies with black holes of millions
to many billions of solar masses. At the moment of maximum compression
of the Universe, a massive merger of black holes occurs, which generates
a powerful flash of relict gravitational radiation and generates
a large black hole. The results obtained impose the following strict
condition on all cyclic models of the Universe: these models should contain
a mechanism for the actual removal of at least the largest black hole
formed by the merger of smaller holes during the cosmological cycle.

It is hypothesized that in addition to black holes, a certain number
of neutron stars may survive from the last cycle of the Universe.
This assumption can be verified by analyzing the characteristic features
of the distribution of pulsars and neutron stars by mass and other parameters.

The considered simple model of the evolution of black holes in a cyclic Universe,
as far as the authors know, is the first one described in the literature.
Therefore, the calculations performed and the conclusions drawn from the model
are not final or evidential. They only point to the importance of considering
the cyclic evolution of the black hole population, which formulates
additional criteria for evaluating cyclic cosmologies and challenges
more complex models of the evolution of black holes and neutron stars.
The described model is only a starting point; future models should be
more precise, specific and take into account:

\begin{list}{$\bullet$}{
        \setlength\leftmargin{5mm} \setlength\topsep{2mm}
        \setlength\parsep{0mm} \setlength\itemsep{2mm} }
    \item  mass distribution of the initial population of black holes;
    \item  inhomogeneity of accretion in time during the cycle;
    \item  inhomogeneity of the accreting medium;
    \item  inclusion of neutron stars in the model, taking into account
    their complex evolution;
    \item  estimation of the influence of the magnetic field and all other possible
    factors not taken into account in the model under discussion.

\end{list}

\subsection{Acknowledgments}
The authors are grateful to A.~Vasilkov, A.~Bogomazov, J.~Mather,
as well as to anonymous reviewers for useful discussions and comments,
and L.~B.~Potapova for help in preparing the work.

\section*{CONFLICT OF INTEREST}
The authors declare no conflicts of interest.

\section*{APPENDIX. OSCILLATING MODEL OF THE UNIVERSE WITH A VARIABLE GRAVITATIONAL MASS}
The results of the LIGO group showed that about $5\%$ of the mass
of merging black holes is converted into gravitational radiation.
The Nobel laureate \citet{Anderson2018} noted that this requires
the construction of a cosmological model with a variable gravitational mass.
Indeed, there is an interpretation of GR that goes back to the works
of Einstein, Eddington, and Schrodinger (see references in the article
\citet{Gorkavyi2016}), according to which gravitational waves
should not be included in the sources of the gravitational field,
therefore, the transition of part of the mass of the Universe
to gravitational radiation reduces its total gravitational mass.
This is exactly what happens when the observed population
of black holes merges during the compression of the Universe
and dumps a significant part of its mass into gravitational waves,
which contribute to the background of the relict gravitational radiation.
Consider a model of a cyclic Universe with a variable gravitational mass.
\citet{Kutchera2003} obtained a modified Schwarzschild metric
for the variable gravitational mass of a fireball in
the weak field approximation:
\begin{equation}
    \begin{array}{rcl}

d{s}^2\!\!&\!\!=\!\!&\!\![1\!-\!b(t,r)]{c^2}d{t}^{2}\!-\![1\!+\!b(t,r)](dx^2\!+\!dy^2\!+\!dz^2),\\

\end{array}
\end{equation}
where $b(t,r)=2GM(t-r/c)/(rc^2)$.

\noindent \citet{Kutchera2003} concluded that a decrease in the gravitational mass
of a fireball generates a monopole gravitational wave. Note that
Birkhoff's theorem, which is often referred to as an argument against
the existence of monopole gravitational waves, is not applicable
to systems with gravitational radiation that do not have spherical symmetry.
In \citet{Gorkavyi2016}, a system consisting of many gravitational wave
emitters is considered, and a similar space-time metric for a variable mass
is independently obtained. Consider a quasi-spherical system
in which gravitational radiation is generated, for example,
during the merger of black holes. For the weak external gravitational field
of this system, we write:
${g}_{{\mu}{\nu}}={\eta}_{{\mu}{\nu}}+{h}_{{\mu}{\nu}}$,
where  $\eta_{{\mu} {\nu}}$ is the Minkowski tensor
and $\eta_{{\mu} {\nu}} \gg h_{{\mu} {\nu}}$. Let's write down
the Einstein equation for the weak field in the known form
(see references in \citet{Gorkavyi2016}):
\begin{equation}
\ \Big({\nabla}^2-{\dfrac{\partial^2}{{c^2}{\partial {t^2}}}}\Big ){h}_{{\mu}{\nu}}=-{\dfrac{16{\pi}G}{c^4}}\Big({T}_{{\mu}{\nu}}-{\dfrac{1}{2}}{\eta}_{{\mu}{\nu}}{T}^{\lambda}_{\lambda}\Big).
\end{equation}

The solution of the wave equation (12) is the delayed potential,
which includes a variable gravitational mass.
The zero component of the metric tensor will take the form
(\citet{Kutchera2003}, \citet{Gorkavyi2016}):
\begin{equation}
\ {g}_{00}=-\Big[1-\dfrac{2GM(t-r/c)}{rc^2}\Big].
\end{equation}

In the paper \citet{Gorkavyi2016} for the modified Schwarzschild metric (11), (13),
the gravitational acceleration was calculated and it was indicated
that the mass merger of black holes at the last stage of the collapse
of the Universe can generate a repulsive force that exceeds
the gravitational attraction. We describe the variable mass of the Universe
with the following function: \mbox{$M={M_0}{e^{-{\alpha}(t-r/c)}}$.}
Assuming a weak gravitational field and low velocities, we get an expression
for the gravitational acceleration:
\begin{equation}
\ F \approx {\dfrac{c^2}{2}}{\dfrac{\partial{{g}_{00}}}{\partial{r}}}= -{\dfrac{GM}{r^2}}+{\dfrac{\alpha}{c}}{\dfrac{GM}{r}}.
\end{equation}

The first term characterizes the Newtonian attraction.  For $\alpha > 0$,
the second term describes ``anti-gravity'', and in the case of $\alpha < 0$---``hypergravity''. To clarify the physical meaning of equation (14), we can
write the gravitational acceleration in terms of the quasi-Newtonian potential
$\phi$: \mbox{$\ F =-\dfrac{{\partial}{\phi}}{\partial{r}}$,}
where $\ {\phi}=-{\dfrac{GM(t-r/c)}{r}}$  (this equation was published by Gorkavyi
in 2003\, (see the reference in \citet{Gorkavyi2016})). After differentiating
this potential, the expression (14) is obtained.  Attraction is analogous
to the motion of the balls along the inclined surface of the potential funnel,
and anti-gravity corresponds to the scattering of the balls if
the potential forms a peak, and not a funnel. An analog of this antigravity
has already been investigated earlier: a similar repulsive force
is responsible for the escape of a black hole formed by the merger
of black holes of unequal masses \citep{Rezzolla2010}. The resulting
gravitational potential is not directed inwards, but outwards, and it throws
the new hole away from the point of fusion, even if it was located in the center
of gravity of the system. It is logical to assume that the second term
of equation (14) is responsible for the Big Bang mechanism.
Consider the conditions under which antigravity will be stronger
than gravitational attraction. Let a system with mass $M$ and radius $R$ decrease
its mass due to its transformation into gravitational waves. From equation (14),
it is not difficult to obtain the condition for the dominance
of antigravity over attraction:
\begin{equation}
\ -{\dfrac{dE}{dt}} > \dfrac{Mc^3}{R}.
\end{equation}
where $E$ is the energy of the gravitational matter. Although this condition
was formally obtained for the case of weak fields, it can be expected that
it will be satisfied for any fields, because a similar formula is obtained
from quasi-Newtonian
calculations, which are not subject to any restrictions \citep{Gorkavyi2016}.
To estimate the radiation of a system of gravitational waves, you can use
the well-known expression for the radiation power of gravitational waves:
\begin{equation}
\ -{\dfrac{dE}{dt}} = S \Big(\dfrac{GM}{Rc^2}\Big)^5 \dfrac{c^5}{G}.
\end{equation}

Here we have introduced the non-sphericity parameter $S$, which is zero
for a perfectly spherical system and of the order of $1$ for a binary star system.
Substituting (16) into (15), we obtain the condition under which antigravity
begins to dominate:
\begin{equation}
S \Big(\dfrac{GM}{Rc^2}\Big)^5 \dfrac{c^5}{G} > \dfrac{Mc^3}{R}.
\end{equation}

It is easy to see that this condition coincides to the accuracy
of the coefficient $S$ with the condition of being inside a black hole:

\begin{equation}
 S \Big(\dfrac{GM}{Rc^2}\Big)^4 > 1.
\end{equation}

\citet{Poplawski2016} develops an elegant model of a cyclical Universe
pulsating inside a huge black hole. Within the framework of such a model,
condition (18) should be satisfied, unless the introduced nonsphericity
parameter $S$ is abnormally small. As shown in many works, gravitational
collapse leads to an increase in nonsphericity. This can also be formulated
in Newtonian language: small fluctuations of the surface of the collapsing
ball will be unstable due to stretching by tidal forces, which grow as
$1/r^3$, that is, faster than gravitational forces: $1/r^2$. When the Universe
is compressed to several light years, its radius is reduced by 10 orders of magnitude,
increasing the right side of (18) by 40 orders of magnitude, so it can be assumed
that at any realistic degree of nonsphericity, condition (18) will be satisfied.
This removes the general singularity problem: any system within
the Schwarzschild radius evaporates into gravitational waves and generates
powerful repulsive forces before reaching the singularity. Note
that this does not contradict the Hawking-Penrose theorems,
which are not applicable to systems with antigravity and with
a positive cosmological constant \citep{Hawking2003}. Thus, the Big Bang
is an expansion of the fireball of the compressed Universe, which got into
the conditions of strong antigravity, which arose during the Big Compression.

As shown in \citet{Gorkavyi2018a}, the problem of dark energy
also finds its logical solution in the cosmological model with
a variable gravitational mass. During the expansion phase of the Universe,
the merging of black holes and the transition of their mass
into gravitational radiation becomes a rare event, and black holes
begin to grow, absorbing the background gravitational radiation.
The biggest black hole (BBH) formed during the collapse of the Universe,
grows faster than all. Let us study the influence of the gravitational field
of the growing BBH on the expansion of the Universe.
We derive the modified Friedman equations for a metric with variable mass
in the comoving coordinates $x_{\ast},y_{\ast},z_{\ast}$:
\begin{equation}
d{s}^2={c^2}d{t}^{2}-a^2(t,r)[1+b(t,r)](dx^2_{\ast}+dy^2_{\ast}+dz^2_{\ast}),
\end{equation}
where $b(t,r)=2GM(t,r)/(rc^2)$ is a known function, and $a(t,r)$ is an unknown
scale factor. For $\mid\alpha\mid \gg c/r$, the dependence of $a(t,r)$
on spatial coordinates is significantly weaker than that of the function
$b(t,r)$ \citep{Gorkavyi2018a}. The paper \citet{Gorkavyi2018a} considered
the more general case of metric (19), which takes into account
the weak inhomogeneity of time, and it was shown that the modified
Friedman equation includes perturbations $b(t,r)$ only from
the spatial part of the metric. For the case of a weak gravitational field
and $b(t,r) \ll 1$, we obtain the first Friedman equation in the form:
\begin{equation}
\Big(\dfrac{\dot{a}}{a}\Big)^{2}=\dfrac{{\Lambda(t,r)}c^{2}}{3}+\dfrac{8{\pi}G{\rho}}{3},
\end{equation}
where the cosmological function $\Lambda(t,r)$ is expressed as follows:
\begin{equation}
    \begin{array}{rcl}
\Lambda(t,r)&=&\dfrac{1}{2g_{11}g_{22}}\dfrac{\partial^2 g_{22}}{\partial x_{\ast}^2}
+\dfrac{1}{2g_{11}g_{33}}\dfrac{\partial^2 g_{33}}{\partial x_{\ast}^2}\\
&+&\dfrac{1}{2g_{11}g_{22}}\dfrac{\partial^2 g_{11}}{\partial y_{\ast}^2}
+\dfrac{1}{2g_{22}g_{33}}\dfrac{\partial^2 g_{33}}{\partial y_{\ast}^2}\\&+&
\dfrac{1}{2g_{11}g_{33}}\dfrac{\partial^2 g_{11}}{\partial z_{\ast}^2}+
\dfrac{1}{2g_{22}g_{33}}\dfrac{\partial^2 g_{22}}{\partial z_{\ast}^2},
\end{array}
\end{equation}
or
\begin{equation}
\Lambda(t,r)=\Big(\dfrac{\partial^2 b}{\partial x^2}+\dfrac{\partial^2 b}{\partial y^2}+\dfrac{\partial^2 b}{\partial z^2}\Big),
\end{equation}
where $x,y,z$ are the physical coordinates. From equation (22)
for $\mid\alpha\mid \gg c/r$ we get
\begin{equation}
    \begin{array}{rcl}
\Lambda(t,r)&\approx &\dfrac{\alpha^{2}}{c^{2}}b(t,r)=\dfrac{\alpha^{2}}{c^{2}}\dfrac{r_{0}}{r}\\
&\approx&
 0.7 \times 10^{-56}{(\alpha T)}^{2}\dfrac{r_{0}}{r}\text{[\rm cm$^{-2}$]} 
\end{array}
\end{equation}
where~$r_{0}=\dfrac{2GM(t,r)}{rc^{2}}$ is the Schwarzschild radius,
\mbox{$T \approx 4 \times 10^{17}$s} is the cosmological time.
$\Lambda(t,r)$ is equal to the observed value of the cosmological constant
$1.1\times10^{-56} \rm{cm^{-2}}$ if ${(\alpha T)}^{2} \dfrac{r_{0}}{r}=1.6$
\citep{Gorkavyi2016}. For example, the latter is true if the dimensionless parameter
\mbox{$\dfrac{r_{0}}{r}=0.016$} and \mbox{$\mid\alpha\mid T=10$}.
The existence of relict gravitational waves is not in doubt
(they are most intensively generated when black holes merge with each
compression of the Universe), but the level of their energy is unknown.
Their total energy was usually limited by the condition that they did not make
a noticeable contribution to the total gravitational density of the Universe.
From the point of view of Einstein-Eddington-Schrodinger, gravitational radiation
does not contribute to the gravitational mass of the Universe,
so this restriction becomes invalid. In the article \citet{Gorkavyi2018b},
the density of the medium of gravitational waves $\rho_{\rm{GW}}$ was determined,
with the absorption of which the BBH increases at a rate sufficient
for the observed value of the cosmological constant:

\begin{equation}
\dot{M}=-\alpha M=27\pi{r_{0}^2}c\rho_{\rm{GW}}.
\end{equation}
Then we get
\begin{equation}
\rho_{\rm{GW}}=\dfrac{\mid\alpha\mid c^3}{108 \pi G^2 M}\sim 10^{-28} \text{\rm {g\,cm$^{-3}$}} 
\end{equation}
where ${\mid\alpha\mid T}=10$ and the BBH mass $M=6\times10^{54}$~g (this BBH mass can be
calculated from the condition \mbox{$\dfrac{r_{0}}{r}\approx0.02$} \citep{Gorkavyi2016}).
If we take ${\mid\alpha\mid T}=100$, then $\dfrac{r_{0}}{r}\approx 2\times10^{-4}$,
the BBH mass $M=6\times10^{52}$~g,
and $\rho_{\rm{GW}}\sim 10^{-25}$~\rm{g\,cm${^{-3}}$.
The perturbation of the gravitational field from BBH, which is now observed
in the form of accelerated recession of galaxies, arose about 13 billion years ago,
therefore, during this time, the energy density of the background
of gravitational radiation should decrease by several orders of magnitude.
Consider the case when a term with a cosmological function dominates over
a term with an average density. For an exponential change in the BBH mass, we obtain
\mbox{$\dot{\Lambda}(t,r)=-\alpha \Lambda(t,r)$.}
Then the second Friedman equation will take the form:
\begin{equation}
\dfrac{\ddot{a}}{a}\approx-\dfrac{\alpha}{2}\sqrt{\dfrac{\Lambda(t,r)c^{2}}{3}}.
\end{equation}
Hence it follows that the acceleration of the observable part of the Universe
$\dfrac{\ddot{a}}{a}>0$ is caused by the effect of hypergravity at $\alpha < 0$.
Hypergravity from the growing BBH (or the population of the largest holes)
stretches the observed set of galaxies so that, from the point of view
of a local observer, the galaxies are moving away from each other
with acceleration. A number of observations indicate an increase in the constant
$H_0$ with time \citep{Riess2020}. This presents a problem for theories
that tie the cosmological constant to the properties of the vacuum.
In the model under consideration, the cosmological constant is a function
of the BBH mass and the density of background gravitational waves
$\rho_{\rm{GW}}$, so the growth of  $H_0$ is admissible.

Let us consider the problem of the transition of the expanding Universe
to contraction. The fact that the observed local average density of matter
is less than the critical density required to close the Universe
does not mean that the Universe is open. If BBH grows, absorbing not only
ordinary matter and radiation, but also background gravitational waves,
then the total gravitational mass of the Universe will also increase.
BBH can expand and include all observed galaxies and stars.
For a resting external observer, the signature of the Schwarzschild metric
inside a black hole ``deteriorates'', which is usually corrected
by a change (redesignation) of the spatial and temporal axes,
leading to exotic dynamics of bodies inside black holes. On the one hand,
this topic has not been studied enough, and other options for correcting
the signature are possible.
On the other hand, the dynamics of a freely falling observer obeys
the principle of equivalence, according to which such a local observer
cannot distinguish his fall from rest. Therefore, a freely falling observer
easily crosses the surface of a black hole in its own reference frame
(from the point of view of an external stationary observer, this is impossible,
because time stops at the boundary of a black hole). From the point of view
of local observers, all their galaxies will freely cross the surface
of a large black hole and will exist and move inside it,
following the laws of General Relativity. This permits the consideration
of cosmology inside a black hole, as, for example, Poplawski does \citep{Poplawski2016}.
Thus, the growth of BBH and the absorption of all matter in the Universe by it
solves the problem of stopping the Universe and its subsequent collapse.
It is well known that the average density required to close the Universe
is the same as the density of a black hole of size of the Universe.
If the Universe is closed, then space should have a slight positive curvature.
Indeed, from the data from the WMAP and Planck satellites, the positive curvature
of the Universe space is established with a confidence of more than
$99\%$ \citep{Valentino2019}. Consider the entropy growth problem,
which is difficult for all cyclic cosmologies. Black holes have huge
entropy \citep{Egan2010}:
\begin{equation}
\ Entropy=\dfrac{4\pi Gk}{\hbar c}{M^2},
\end{equation}
where $k$ is the Boltzmann constant. It is known that the main entropy
of the Universe is contained in the largest black holes \citep{Egan2010}.
From this formula, we obtain the following estimate of the entropy value
for the population of black holes: $Entropy\sim10^{77}(M_{\odot})^2N[k]$.
Entropy for a population of black holes with the same hole mass
of 30 solar masses (assuming they make up all dark matter):
\mbox{$M=30M_{\odot};$} \mbox{$N=10^{22}$;} $Entropy\sim10^{102}[k]$.
The entropy of one BBH is 20 orders of magnitude higher:
\mbox{$M=5\times 10^{22}M_{\odot}$;} $N=1; Entropy\sim2.5\times10^{122}[k]$.
Thus, practically all the entropy of the Universe will be contained
in the BBH that arose in this cycle, so the entropy problem can be formulated
as follows: what to do with the BBH that arises at each cycle of the Universe?
Obviously, the only way to get rid of BBH is to get into it and start
a new cosmological cycle inside, which is what our scenario suggests.
In this case, the entropy problem is solved automatically:
the Bekenstein--Hawking entropy is determined only for an outside observer
who measures a certain thermal temperature of a black hole.
This temperature, and hence the entropy of the black hole, is not determined
for the internal observer. Thus, a set of galaxies, getting inside a black hole,
find themselves in a system whose entropy is twenty orders of magnitude less
than the entropy of the Universe at the stage of expansion.
This sharp decrease in entropy also manifests itself in the dynamics of matter
and radiation: at the stage of expansion of the Universe,
the density of matter dropped, as did the temperature
of the background radiation, and after entering the big black hole,
the density of matter begins to grow, as does the temperature of radiation.
The surface of the big black hole, having absorbed all matter
and background radiation of the Universe, goes to the periphery,
and inside this huge hole the observed part of the Universe begins
a new cycle under conditions of reduced entropy. Formally,
the entropy of the Universe as a whole, taking into account the surface
of the outer outer shell of a black hole (slowly increasing or ``multilayered''),
grows with each cycle, but in practice, the observed set of galaxies
enters a new cycle, dropping entropy to a minimum.
Thus, the Universe turns out to be a closed system or an object
with variable gravitational mass, pulsating inside a huge black hole.
A similar model has been considered not only by \citet{Poplawski2016},
but also by other authors (see references in \citet{Gorkavyi2018a}).
This cosmological model, in addition to baryons, gravitational
and electromagnetic radiation, contains an extensive population
of black holes, the evolution of which is the subject of study
in the main text of this article.




\begin{thebibliography}{99}

    \bibitem[Abbott et~al.(2016)]{Abbot2016}
    B.~P. Abbott, R.~Abbott, T.~D. Abbott, et~al., Phys. Rev. Lett. \textbf{116},
    061102 (2016).

    \bibitem[{Abbott} et~al.(2017)]{Abbot2017}
    B.~P. {Abbott}, R.~{Abbott}, T.~D. {Abbott}, et~al., \apjl \textbf{848}~(2),
    L13 (2017).

        \bibitem[{Event Horizon Telescope Collaboration} et~al.(2019)]{Akiyama2019}
    K.~{Akiyama} et~al. (Event Horizon Telescope Collaboration)
    \apjl \textbf{875}~(1), L5 (2019).

        \bibitem[{Planck Collaboration} et~al.(2020)]{Akrami2019}
     Y.~{Akrami} et~al. (Planck Collaboration), \aap \textbf{641},
    A7 (2020).

    \bibitem[{Anderson}(2018)]{Anderson2018}
    P.~W. {Anderson}, arXiv:1804.11186 (2018).

    \bibitem[{Arzamasskiy} et~al.(2017)]{Arzamasskiy2017}
    L.~I. {Arzamasskiy}, V.~S. {Beskin}, and K.~K. {Pirov}, \mnras
    \textbf{466}~(2), 2325 (2017).

    \bibitem[{Ba{\~n}ados} et~al.(2018)]{Banados2017}
    E.~{Ba{\~n}ados}, B.~P. {Venemans}, C.~{Mazzucchelli}, et~al., \nat
    \textbf{553}~(7689), 473 (2018).

    \bibitem[{Barkana} and {Loeb}(2001)]{Barkana2001}
    R.~{Barkana} and A.~{Loeb}, \physrep \textbf{349}~(2), 125 (2001).

    \bibitem[{Bekenstein}(1973)]{Bekenstein1973}
    J.~D. {Bekenstein}, \prd \textbf{7}~(8), 2333 (1973).

    \bibitem[{Belotsky} et~al.(2019)]{Belotsky2019}
    K.~M. {Belotsky}, V.~I. {Dokuchaev}, Y.~N. {Eroshenko}, et~al.,
     European Phys. J. C \textbf{79}~(3), 246 (2019).

    \bibitem[{Bender} et~al.(2005)]{Bender2005}
    R.~{Bender}, J.~{Kormendy}, G.~{Bower}, et~al., \apj \textbf{631}~(1), 280
    (2005).

    \bibitem[Bird et~al.(2016)]{Bird2016}
    S.~Bird, I.~Cholis, J.~B. Mu\~noz, et~al., Phys. Rev. Lett. \textbf{116},
    201301 (2016).

    \bibitem[{Blinnikov} et~al.(1984)]{Blinnikov1984}
    S.~I. {Blinnikov}, I.~D. {Novikov}, T.~V. {Perevodchikova}, and A.~G.
    {Polnarev}, Sov. Astron. Lett. \textbf{10}, 177 (1984).

    \bibitem[{Bondi}(1952)]{Bondi1952}
    H.~{Bondi}, \mnras \textbf{112}, 195 (1952).

    \bibitem[{Boyarchuk} and {Tanzi}(1993)]{Boyarchuk1993}
    A.~A. {Boyarchuk} and E.~G. {Tanzi}, \memsai \textbf{64}, 263 (1993).

    \bibitem[{Brandenberger} and {Peter}(2017)]{Brandenberger2017}
    R.~{Brandenberger} and P.~{Peter}, Foundations of Physics \textbf{47}~(6), 797
    (2017).

    \bibitem[{Bullock} and {Boylan-Kolchin}(2017)]{Bullock2017}
    J.~S. {Bullock} and M.~{Boylan-Kolchin}, \araa \textbf{55}~(1), 343 (2017).

    \bibitem[{Carr} et~al.(2020)]{Carr2020}
    B.~{Carr}, K.~{Kohri}, Y.~{Sendouda}, and J.~{Yokoyama}, arXiv:2002.12778 (2020).

    \bibitem[{Carr} and {Silk}(2018)]{Carr2018}
    B.~{Carr} and J.~{Silk}, \mnras \textbf{478}~(3), 3756 (2018).

    \bibitem[Cherepashchuk(2014)]{Cherepashchuk2014}
    A.~M. Cherepashchuk, Physics-Uspekhi  \textbf{57}~(4), 359 (2014).

    \bibitem[Cherepashchuk(2016)]{Cherepashchuk2016}
    A.~M. Cherepashchuk, Physics-Uspekhi  \textbf{59}~(7), 702 (2016).

    \bibitem[{Clesse} and {Garc{\'\i}a-Bellido}(2017)]{Clesse2017}
    S.~{Clesse} and J.~{Garc{\'\i}a-Bellido}, Physics of the Dark Universe
    \textbf{15}, 142 (2017).

    \bibitem[{Clesse} and {Garc{\'\i}a-Bellido}(2018)]{Clesse2018}
    S.~{Clesse} and J.~{Garc{\'\i}a-Bellido}, Physics of the Dark Universe
    \textbf{22}, 137 (2018).

    \bibitem[{Clifton} et~al.(2017)]{Clifton2017}
    T.~{Clifton}, B.~{Carr}, and A.~{Coley}, Classical and Quantum Gravity
    \textbf{34}~(13), 135005 (2017).

    \bibitem[{Cordes} and {Shannon}(2008)]{Cordes2008}
    J.~M. {Cordes} and R.~M. {Shannon}, \apj \textbf{682}~(2), 1152 (2008).

    \bibitem[{Di Valentino} et~al.(2020)]{Valentino2019}
    E.~{Di Valentino}, A.~{Melchiorri}, and J.~{Silk}, Nature Astronomy \textbf{4},
    196 (2020).

    \bibitem[{Dicke} et~al.(1965)]{Dicke1965}
    R.~H. {Dicke}, P.~J.~E. {Peebles}, P.~G. {Roll}, and D.~T. {Wilkinson}, \apj
    \textbf{142}, 414 (1965).

    \bibitem[{Dolgov} and {Postnov}(2017)]{Dolgov2017}
    A.~{Dolgov} and K.~{Postnov}, \jcap \textbf{2017}~(4), 036 (2017).

    \bibitem[Dolgov(2018)]{Dolgov2018}
    A.~D. Dolgov, Physics-Uspekhi  \textbf{61}~(2), 115 (2018).

    \bibitem[{Egan} and {Lineweaver}(2010)]{Egan2010}
    C.~A. {Egan} and C.~H. {Lineweaver}, \apj \textbf{710}~(2), 1825 (2010).

    \bibitem[Fridman and Gorkavyi(1999)]{Gorkavyi1994}
    A.~M. Fridman and N.~N. Gorkavyi, \emph{Physics of Planetary Rings. Celestial
        Mechanics of a Continuous Media} (Springer-Verlag, Berlin, 1999).

    \bibitem[Garcia-Bellido(2018)]{Garsia2018}
    J.~Garcia-Bellido, {in {\it Exploring the Dark Side of the Universe}},
    PoS(EDSU2018)042 (https://pos.sissa.it/335/042/) (2018).

    \bibitem[{Gerssen} et~al.(2002)]{Marvel2002}
    J.~{Gerssen}, R.~P. {van der Marel}, K.~{Gebhardt}, et~al., \aj
    \textbf{124}~(6), 3270 (2002).

    \bibitem[{Gorkavyi} and {Vasilkov}(2016)]{Gorkavyi2016}
    N.~{Gorkavyi} and A.~{Vasilkov}, \mnras \textbf{461}~(3), 2929 (2016).

    \bibitem[{Gorkavyi} and {Vasilkov}(2018)]{Gorkavyi2018a}
    N.~{Gorkavyi} and A.~{Vasilkov}, \mnras \textbf{476}~(1), 1384 (2018).

    \bibitem[{Gorkavyi} et~al.(2018)]{Gorkavyi2018b}
    N.~{Gorkavyi}, A.~{Vasilkov}, and J.~{Mather}, in \emph{Exploring the Dark Side
        of the Universe}, PoS(EDSU2018)039  (https://pos.sissa.it/335/039/) (2018).

    \bibitem[Gorkavyi(2021)]{Gorkavyi2020}
    N.~N. Gorkavyi, New~Astronomy \textbf{89} (2021, in press).

    \bibitem[{Hawking}(1975)]{Hawking1975}
    S.~W. {Hawking}, Commun. Mathematical Physics \textbf{43}~(3), 199
    (1975).

    \bibitem[{Hawking} and {Penrose}(1970)]{Hawking2003}
    S.~W. {Hawking} and R.~{Penrose}, Proc. Royal Soc. London Ser. A \textbf{314}~(1519), 529 (1970).

    \bibitem[{Johnston} and {Karastergiou}(2017)]{Jonston2017}
    S.~{Johnston} and A.~{Karastergiou}, \mnras \textbf{467}~(3), 3493 (2017).

    \bibitem[Kardashev et~al.(2000)]{Kardashev2000}
    N.~Kardashev, V.~Andreyanov, V.~Buyakas,  et~al., Tr.
        Fizicheskogo inst. im. P.~N.~Lebedeva \textbf{228},
    pp. 112 (2000).

    \bibitem[{Kashlinsky}(2016)]{Kashlinsky2016}
    A.~{Kashlinsky}, \apjl \textbf{823}~(2), L25 (2016).

    \bibitem[Kippenhahn(1987)]{Kippenhan1990}
    R.~Kippenhahn, \emph{Hundert Milliarden Sonnen. Geburt, Leben and Tod der
        Sterne} (Piper, Munchen Zurich, 1987).

    \bibitem[{Klypin} et~al.(1999)]{Klypin1999}
    A.~{Klypin}, A.~V. {Kravtsov}, O.~{Valenzuela}, and F.~{Prada}, \apj
    \textbf{522}~(1), 82 (1999).

    \bibitem[{Kutschera}(2003)]{Kutchera2003}
    M.~{Kutschera}, \mnras \textbf{345}~(1), L1 (2003).

    \bibitem[Landau and Lifshitz(1980)]{Landau1987}
    L.~D. Landau and E.~M. Lifshitz, \emph{Statistical Physics}, vol.~5, 3rd ed.
    (Pergaman Press, Oxford, 1980).

    \bibitem[{Li}(2006)]{Li2006}
    X.-D. {Li}, \apjl \textbf{646}~(2), L139 (2006).

    \bibitem[Lorimer and Kramer(2005)]{Lorimer2005}
    D.~R. Lorimer and M.~Kramer, \emph{Handbook of pulsar astronomy} (Cambridge
    Univ. Press, Cambridge 2005).

    \bibitem[{Ma} et~al.(2007)]{Ma2007}
    J.~{Ma}, R.~{de Grijs}, D.~{Chen}, et~al., \mnras \textbf{376}~(4), 1621
    (2007).

    \bibitem[{Malov}(2001)]{Malov2001}
    I.~F. {Malov}, \arep \textbf{45}~(5), 389 (2001).

    \bibitem[{McLaughlin} et~al.(2006)]{McLaughlin2006}
    M.~A. {McLaughlin}, A.~G. {Lyne}, D.~R. {Lorimer}, et~al., \nat
    \textbf{439}~(7078), 817 (2006).

    \bibitem[{Narayan} and {Ostriker}(1990)]{Narayan1990}
    R.~{Narayan} and J.~P. {Ostriker}, \apj \textbf{352}, 222 (1990).

    \bibitem[{Novello} and {Bergliaffa}(2008)]{Novello2008}
    M.~{Novello} and S.~E.~P. {Bergliaffa}, \physrep \textbf{463}~(4), 127 (2008).

    \bibitem[{{\"O}zel} and {Freire}(2016)]{Ozel2016}
    F.~{{\"O}zel} and P.~{Freire}, \araa \textbf{54}, 401 (2016).

    \bibitem[{{\"O}zel} et~al.(2012)]{Ozel2012}
    F.~{{\"O}zel}, D.~{Psaltis}, R.~{Narayan}, and A.~{Santos Villarreal}, \apj
    \textbf{757}~(1), 55 (2012).

    \bibitem[Patria(1972)]{Patria1972}
    R.~K. Patria, Nature \textbf{240}, 2985 (1972).

    \bibitem[{Payne-Gaposchkin}(1979)]{Payne1979}
    C.~{Payne-Gaposchkin}, \emph{{Stars and clusters}} (Harvard Univ. Press, Cambridge, Mass. 1979).

    \bibitem[{Peebles}(1993)]{Peebles1993}
    P.~J.~E. {Peebles}, \emph{{Principles of Physical Cosmology}} (Princeton
    Univ. Press, Princeton, New Jersey, 1993).

    \bibitem[Penrose(2011)]{Penrose2011}
    R.~Penrose, \emph{Cycles of Time} (Alfred A. Knopf Publ., New York, 2011).


    \bibitem[{Phinney} and {Blandford}(1981)]{Phinney1981}
    E.~S. {Phinney} and R.~D. {Blandford}, \mnras \textbf{194}, 137 (1981).


    \bibitem[{Pop{\l}awski}(2016)]{Poplawski2016}
    N.~{Pop{\l}awski}, \apj \textbf{832}~(2), 96 (2016).

    \bibitem[{Rezzolla} et~al.(2010)]{Rezzolla2010}
    L.~{Rezzolla}, R.~P. {Macedo}, and J.~L. {Jaramillo}, Phys. Rev. Lett.
    \textbf{104}~(22), 221101 (2010).

    \bibitem[{Riess}(2020)]{Riess2020}
    A.~G. {Riess}, Nature Rev. Phys. \textbf{2}~(1), 10 (2020).

    \bibitem[Smith(1977)]{Smith1979}
    F.~G. Smith, \emph{Pulsars} (Cambridge University Press, London, New York,
    1977).

    \bibitem[{Steinhardt} and {Turok}(2002)]{Steinhardt2002}
    P.~J. {Steinhardt} and N.~{Turok}, Science \textbf{296}~(5572), 1436 (2002).

    \bibitem[{Stuckey}(1994)]{Stuckey1994}
    W.~M. {Stuckey}, Amer. J. Physics \textbf{62}~(9), 788 (1994).

    \bibitem[{Tang} et~al.(2019)]{Tang2019}
    J.-J. {Tang}, T.~{Goto}, Y.~{Ohyama}, et~al., \mnras \textbf{484}~(2), 2575
    (2019).

    \bibitem[{Taylor} et~al.(2015)]{Taylor2015}
    M.~A. {Taylor}, T.~H. {Puzia}, M.~{Gomez}, and K.~A. {Woodley}, \apj
    \textbf{805}~(1), 65 (2015).

    \bibitem[{Vielva}(2010)]{Vielva2010}
    P.~{Vielva}, Advances in Astronomy \textbf{2010}, 592094 (2010).

    \bibitem[{Yakovlev} and {Pethick}(2004)]{Yakovlev2004}
    D.~G. {Yakovlev} and C.~J. {Pethick}, \araa \textbf{42}~(1), 169 (2004).


\end{thebibliography}








\bsp	
\label{lastpage}
\end{document}